\documentclass[onecolumn]{article}
\usepackage{rotating}
\usepackage{setspace}
\usepackage[margin=1.5in]{geometry}
\usepackage[usenames,dvipsnames,svgnames,table]{xcolor}
\usepackage[utf8x]{inputenc}
\usepackage{soul,color}
\soulregister\cite7
\soulregister\ref7
\soulregister\pageref7
\soulregister\citep7
\soulregister\citet7
\usepackage{amssymb,amsmath}
\usepackage{mathrsfs}  
\usepackage{varioref}
\usepackage{wrapfig}
\usepackage{threeparttable}
\usepackage{dcolumn}
\newcolumntype{d}{D{.}{.}{-1}}
\usepackage{nomencl}
\makeglossary
\usepackage{subfigure}
\usepackage{subfigmat}
\usepackage{fancyvrb}
\fvset{fontsize=\footnotesize,xleftmargin=2em}
\usepackage{lettrine}
\usepackage[dvips]{dropping}
\usepackage{multirow}
\usepackage{accents}  
\usepackage{placeins}

\usepackage{hyperref}   
\usepackage{booktabs} 

\DeclareMathAlphabet{\mathcalligra}{T1}{calligra}{m}{n}
\newcommand{\bg}{\mathbf{g}}

\newcommand{\bi}{\mathbf{i}}     
\newcommand{\bj}{\mathbf{j}}
\newcommand{\bk}{\mathbf{k}}

\newcommand{\bq}{\mathbf{q}}      
\newcommand{\br}{\mathbf{r}}

\newcommand{\bu}{\mathbf{u}}

\newcommand{\bx}{\mathbf{x}}     

\newcommand{\bA}{\mathbf{A}}     

\newcommand{\bI}{\mathbf{I}}

\newcommand{\bK}{\mathbf{K}}

\newcommand{\bM}{\mathbf{M}}

\newcommand{\bR}{\mathbf{R}}

\newcommand{\bT}{\mathbf{T}}

\newcommand{\bV}{\mathbf{V}}      

\newcommand{\bzero}{\mathbf{0}}

\newcommand{\om}{\omega}

\newcommand{\bom}{\boldsymbol{\om}}  

\newcommand{\ud}{\mathrm{ d}}
\newcommand{\ddi}{\mathrm{ d}}
\def\pd{\partial}                  
\newcommand{\x}{\! \times \!}           

\newcommand{\mathcalbf}[1]{\boldsymbol{\mathcal{#1}}}  
\newcommand{\mbbm}[1]{{\mbox{\boldmath $ #1 $}}}

\newcommand{\rT}{\!\text{T}}              

\begin{document}

\title{Studies on Coupled Flight Dynamics and Aeroelasticity of a PrandtlPlane Configuration}
\author{Rocco~Bombardieri \\
PhD Candidate, Universidad Carlos III de Madrid \and
Rauno~Cavallaro \\
Faculty Member, Universidad Carlos III de Madrid \and
Rodrigo~Castellanos \\
MSc Student, Universidad Carlos III de Madrid \and
Francesco~Auricchio \\
Visiting MSc Student, Universidad Carlos III de Madrid}


\maketitle


%
\begin{abstract}
The PrandtlPlane aircraft has been recently considered 
as a possible candidate to foster the ambition of a greener aviation. 
Despite the relevant amount of research carried out
in the last years, several aspects of this novel configuration still need
further insight
to pave the way to future applications. %
Among them, the coupled flight-dynamic and aeroelastic response still deserves further investigation, and will be addressed in this work 
%
by means of a dedicated in-house framework. 
%
For the evaluation of the aerodynamic forces, an enhanced Doublet Lattice Method, able to take into account terms typically neglected by classic formulations, is integrated in the framework. 
\par
First, flight-dynamic aspects are considered, showing how effects of interaction between Short Period or Dutch Roll with elastic modes remarkably deteriorate the flying qualities. 
Then, focus is on the aeroelastic stability of the aircraft. 
As observed also in previous literature efforts on this configuration, flutter onset is considerably different when considering the aircraft being free in the air or fixed in space. 
Thanks to the adopted formulation
it is shown how, for this PrandtlPlane, the aerodynamic coupling of elastic and rigid modes has a beneficial effect on flutter onset. 
However, the different modal properties, consequence of the diverse boundary conditions,
when switching from fixed-in-space to free-flying aircraft, also play a relevant role in determining the occurrence of flutter.
Whereas for the longitudinal case both effects are synergistic, contributing to increase flutter speed, for the lateral-directional case the variation in modal properties has a detrimental and dominating effect, leading to a flutter speed well within the flight envelope.
\par
Finally, the work discusses the contributions of the 
additional terms modeled by the enhanced Doublet Lattice Method, showing 
how they induce a considerable effect when modeling the flight dynamics of the flexible aircraft. 
\end{abstract}

\section{Introduction}    
\label{sec:introduction_UFFD}
%
%
%
%


%
%

{C}{urrent} trends in aircraft design push towards lighter and more efficient structures to reduce environmental impact and pave the way for a more sustainable aviation.
As a consequence, interactions between disciplines, previously considered as acting in isolation, are enhanced.
This is the case, for instance, for flight dynamics and aeroelasticity: with lower structural natural frequencies due to more flexible designs, flight-dynamic and aeroelastic behaviors of the aircraft can be strongly coupled; the whole problem should be tackled by a unified approach. 
%
%
Not only are the above mentioned issues exacerbated by more efficient structural design carried out on traditional layouts, but they can be also triggered by unconventional architectures: Flying Wings are the most notable case \cite{Love_Zink_FlyingWing_AIAA2005-1947}.
\par
The relevance of the interaction among flight dynamics and aeroelasticity has been long acknowledged and efforts have been devoted to define models to tackle both disciplines at unified level. 
One of the first contributions on the topic dates back to Milne's paper~\cite{Milne1962},  studying one of the fundamental aspects of flexible flight dynamics, i.e., the choice of the reference frame. 
Following this effort, Etkin~\cite{Etkin_2012} adopted the mean-axes frame (also used in this study) and proposed to model the vehicle elastic deformations as a superposition of normal modes. 
Cavin \textit{et al.}~\cite{Cavin1977} mentioned the issue of the consistency between the mean-axes frame (in which the equations of motion are formally defined) and the frame in which aerodynamic forces are calculated. 
Waszak \textit{et al.}~\cite{Waszak1987} modeled the effects of structural flexibility for a generic family of aircraft within a flight simulator, at NASA Langley VMS simulation facility. 
Complete response data and subjective pilot ratings were investigated. 
Results pointed out degraded aircraft handling qualities when taking into account structural flexibility.
Buttrill \textit{et al.}~\cite{Buttrill1987} presented a mathematical model integrating nonlinear rigid-body flight mechanics and linear aeroelastic dynamics for the study of a flexible aircraft in maneuvering flight. The equations were defined using Lagrangian approach in mean-axes frame and normal modes decomposition approach was used to deal with the vehicle flexibility. 
A nonlinear energy term, accounting for coupling between rigid-body angular velocity and elastic deformations, and usually neglected in traditional formulations on the basis of engineering judgment, was retained and applied to the study of the open-loop response of an F/A-18 model to check the validity of the assumption.  
In a following effort~\cite{Waszak1988}, Waszak \textit{et al.} presented a unified model based on Lagrange's equations and the principle of virtual work to introduce aerodynamic forces. The model was based on the use of mean-axes frame and normal modes decomposition to introduce flexibility of the aircraft. 
Analytical closed-form expressions of Generalized Aerodynamic Forces (GAF) were provided using aerodynamic strip theory (see also ~\cite{Schmidt_FDF_JoA2016}), allowing to observe and get physical insight into effects of aerodynamic parameters variation. 
The method was claimed to be useful in the early design process to explore aeroelastic effects on the vehicle flight-dynamics.
Later on, the model was applied by Schmidt \textit{et al.}~\cite{Schmidt1998} on a large high-speed commercial transport.
\par
Baldelli \textit{et al.}~\cite{Baldelli_JoA2006} presented a method to expand an aeroelastic model to include the classic flight-dynamic equations. 
The work discusses several important aspects, such as the integration of an unsteady panel method (i.e., ZAERO) and the incorporation of aerodynamic data from different sources (e.g. wind-tunnel- and/or flight-test-measured aerodynamic stability derivatives), consistently with the different strategies to model aerodynamic forces, i.e., quasi-steady or unsteady approximations. 
Further discussion about the integration of aerodynamic solvers into a unified  framework can be found in \cite{Kier2009UNIFYINGMA}, proposing a more physical-based rational function approximation, which allows for an easier separation between quasi-steady and unsteady contributions.  
\par
Structural nonlinearities were considered in the work of Changchuan \textit{et al.}~\cite{Changchuanetal_AIAAJ2018}, within a framework able to assess stability properties of a very flexible aircraft, taking into account aeroelastic and flight-dynamic coupling. 
The reference condition, about which small perturbation equations were formulated, was evaluated with a nonlinear trim analysis 
carried out by means of a nonlinear finite element method and a Vortex Lattice Method (VLM). 
The mean-axes reference frame was used to simplify the dynamic equations and to eliminate the inertial coupling between rigid-body motion and vibration modes.
\par
%
%
%
\par
Issues due to coupled flight dynamics and aeroelasticity have been recently observed on several UAVs~\cite{Kotikalpudi_2015_BFF,Schmidt_FDF_JoA2016,Schafer_2018_DLR}, featuring layouts resembling the Flying Wings. 
However, there is another aircraft architecture that seems to be inherently prone to this coupling: the Box Wing~\cite{SDSUteam_7}. 
Box-wing configurations, known also as PrandtlPlanes~\cite{SDSUteam_JWReview}, exhibit a closed wing-system which, when seen frontally, reminds of a box, hence the name.
%
The driving reason for arranging the lifting system in such a way is the reduction of induced drag, as first shown by Prandtl~\cite{Prandtl_1924}.
However, according to the literature, other relevant benefits 
may be gained if a suitable aircraft synthesis is carried out~\cite{Frediani_general_springerII}.
%
\par
The first attempt to perform an aircraft synthesis of a Box Wing was carried out by Lockheed Martin~\cite{Lange_1974}. 
Aeroelastic instabilities, driven by the coupling between rigid-body and elastic modes were observed, even though a more in depth-analysis was not performed to characterize the phenomenon.
After this effort, only few studies were devoted to this configuration, before the work of Frediani's group on the PrandtlPlane, spanning two decades~\cite{Frediani_general_springerII,Karim_2018_Parsifal}.
%
%
To witness the interest of the aerospace community towards this innovative configuration, the recently H2020-funded project PARSIFAL~\cite{Karim_2018_Parsifal} aimed at demonstrating how the payload capacity of present aircraft like the Airbus 320 or Boeing 737 could be raised to the capacity of larger airplanes like A330/B767 by adopting the PrandtlPlane configuration, and, hence, contributing to cut emissions. 
\subsection{Contributions of the present study}    
%
Literature efforts studying aeroelastic behavior of box-wing or PrandtlPlane configurations are limited~\cite{JWchallenge}. 
A few publications considered, to some extent, the interaction between flight dynamics and aeroelasticity and noticed how such coupling is not to be neglected~\cite{SDSUteam_6jour,SDSUteam_8scitech}. 
However, these investigations mostly focused on the aeroelastic side, and never properly considered flight dynamics. 
Likewise, the differences in the aeroelastic behavior considering the free-flying aircraft or the fixed-in-space wing system were never understood in depth.
Moreover the employed aerodynamic model, based on the classic Doublet Lattice Method (DLM) was not necessarily the most suitable one for catching the complex physics of the peculiar layout of the PrandtlPlane configuration: the featured rear wing/fin intersection could potentially present similar issues to the ones observed on T-tails configurations, for which the classic DLM 
is not capable of accurately predicting all the relevant aerodynamic loads \cite{VanZyl-2011}, with a consequent incorrect assessment of the  aeroelastic stability properties. 
%
\par
The aim of this paper is to shed some light on these open issues, to further understand the PrandtlPlane configuration and with the objective of contributing in increasing its Technology Readiness Level, which has consistently advanced with the activities performed within the PARSIFAL project~\cite{Karim_2018_Parsifal,Picch-2020}.  
To this aim, a mathematical model based on the works~\cite{Waszak1988,Baldelli_JoA2006} and integrating an enhanced in-house DLM-method~\cite{VanZyl-2011} in the equations of motion of the flexible aircraft is shown and implemented. 
By means of this formulation both longitudinal and lateral-directional coupled flight dynamic and aeroelasticity of a PrandtlPlane aircraft are investigated.
\par
With respect to flight dynamics, the degradation of flying qualities for the Short Period and Dutch Roll due to the flexibility of the system are quantified and discussed.
%
%
With respect to aeroelasticity, new findings contribute to explain results observed in a few literature efforts, and provide a more precise physical interpretation.
This has been pursued 
by selectively including or excluding aerodynamic coupling effects between rigid and elastic equations governing the dynamics of the flexible aircraft.
%
In particular, for the longitudinal case, it is proven that aerodynamic coupling has a relevant beneficial effect in postponing flutter speed. 
For the lateral-directional case, thanks to this approach, it is demonstrated that the drop in flutter speed, observed for the free-flying configuration, as opposed to the fixed-in-space case, 
is not induced by the rigid/elastic aerodynamic coupling 
but, rather, by the different structural boundary conditions changing the free vibration responses, and hence, the stability properties.
%
%
\par
Finally, the adoption of the enhanced DLM within the formulation is discussed.
On the one hand, it proves to play a key role in correctly predicting the stability of the Dutch Roll, exacerbating 
structural flexibility effects on the flight dynamic response.
On the other hand, flutter point is not affected to a noticeable extent, suggesting that, in this case, if aeroelastic aspects are to be studied, the terms traditionally considered in the DLM might retain the relevant physics for an accurate flutter prediction.
%
%
\par
These new findings contribute to a deeper understanding of coupled aeroelastic and flight-dynamic response of this novel configuration, and could support the transition from early to later design stages. 
%
%
\par
%
%
%

 \section{Theoretical background: Flight Dynamics of a Flexible Aircraft}   
\label{s:FDDA}
This section outlines the procedure that yields the equations governing the dynamics of a flexible aircraft. 
One of the key points is the choice of the \emph{mean-axes} reference system, which allows to decouple the inertial set of equations due to the motion of the ``whole'' aircraft from the ones relative to its elastic deformations.
A second key point is the integration of 
an enhanced Doublet Lattice Method (DLM) for the evaluation of the aerodynamic forces.
%
%
\subsection{Equations of Motion}   %
%
Equations of motion are obtained using Lagrange's approach:
%
\begin{equation}
\begin{aligned}
\frac{\ud}{\ud t}\left(\frac{\pd T}{\pd \dot{\bq}}\right)-\frac{\pd T}{\pd {\bq}}+\frac{\pd U}{\pd \bq}=\mathbf{Q}= \frac{\pd \left(\delta \mathcal{W}\right)}{\pd \left(\delta{\bq}\right)}^T
\end{aligned}
\label{eq:lagrange}
\end{equation}
where:
\begin{itemize}
\item $T$ is the total kinetic energy of the system;
\item $U$ is the potential energy of the system, including the strain energy of the elastically-deformed body;
\item $\bq$ is the vector of generalized coordinates used to describe the system;
\item $\mathbf{Q}$ is the vector of the GAF acting on the system;
\item $\delta \bq$ is the vector of generalized virtual displacements; 
\item $\delta \mathcal{W}$ is the virtual work of the aerodynamic forces acting on the vehicle.
\end{itemize}
%
%
\subsubsection{Reference systems}        %
\begin{figure}[htb]
\centering
\includegraphics[width=0.8\textwidth]{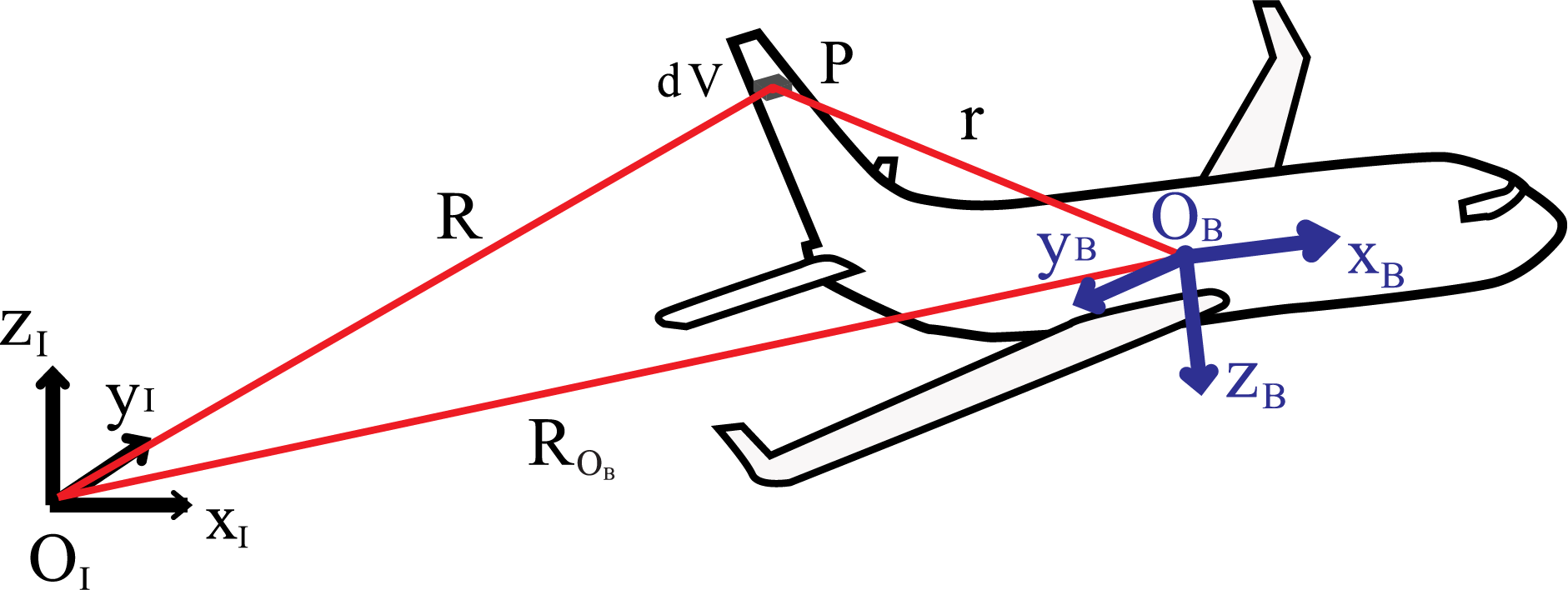}
\caption{Inertial and body frames and position vectors.}
  \label{f:inertial_vehicle_frames}
\end{figure}

Two different reference frames are considered: an inertial one $\left( \Sigma_I \right) $, with origin $O_I$, and a body (vehicle-fixed) one $\left( \Sigma_B \right)$, with origin $O_B$.
The inertial position $\bR$ of a material point P (of infinitesimal volume $\ddi V$) can be expressed as the sum of the position of the body frame origin $\bR_{O_B}$, and the relative position of P in the body frame $\br$, as depicted in Fig.~\ref{f:inertial_vehicle_frames}: 
%
\begin{equation}
\begin{aligned}
\bR=\bR_{O_B}+\br\end{aligned}
\label{eq:pos}
\end{equation}
%
%
%
\subsubsection{Elastic displacements}  %
The relative position $\br$ of a 
point P can be decomposed into its position in the rigid configuration $\br_{RB}$ and its elastic displacement $\bu$: 
%
\begin{equation}
\br = \br_{RB} + \bu
\end{equation}
%
Notice that, with such decomposition, the assumption that the Center of Mass (CoM) does not change when the aircraft is in its rigid configuration has been implicitly done. Hence, $\bu$ is only driven by elastic deformations of the structure. 
\par
Here and for the rest of the paper, the hypothesis of ``linear'' behavior of the structure is assumed; more specifically, small elastic strains and strain rates are assumed. 
%
\par
As customary in structural dynamics, elastic displacements are approximated as a superposition of the first $n$ normal modes:
%
%
\begin{equation}
\bu(x,y,z,t) \approx \sum_{i=1}^{n_E} \boldsymbol{\nu}_i\left(x,y,z\right){\eta}_{{E}_i}\left(t\right)
\label{eq:modal_decomp}
\end{equation}
%
%
where $\eta_{E_i}$ is the modal coordinate relative to the i-th elastic normal mode and $\boldsymbol{\nu}_i\left(x,y,z\right)$ represents the i-th mode shape. A finite number of modes $n_E$ is used, so the model is based on a truncated-mode description.
%
\subsubsection{Generalized coordinates and kinematic equations}  %
%
The inertial velocity of the vehicle is 
defined by the rate of displacement of its center of mass, hence: 
%
\begin{equation}
\begin{aligned}
\bV_{O_B} = \left.\frac{\ud \bR_{O_B}}{\ud t}\right|_I = \dot{R}_{O_{B_X}} \bi_I+ \dot{R}_{O_{B_Y}} \bj_I+ \dot{R}_{O_{B_Z}} \bk_I
\end{aligned}
\label{eq:intertial_V}
\end{equation}
%
with $\bi_I$, $\bj_I$, $\bk_I$ the unit vectors in the three directions of the inertial Frame $\Sigma_I$, ${R}_{O_{B_X}}$, ${R}_{O_{B_Y}}$, ${R}_{O_{B_Z}}$ the components of the inertial position $\bR_{O_B}$ and $\left. \frac{\ud }{\ud t} \right|_I$ is the derivative as measured in the inertial frame.
\par
The inertial orientation of the body-reference frame is described by Euler angles $\Phi$, $\Psi$ and $\Theta$, according to the Tait-Bryan formalism.
The components of the inertial position and the Euler angles describe the rigid-body Degrees of Freedom (DoFs) of the vehicle. 
The set of generalized coordinates of the system is finally obtained:
%
\begin{equation}
\begin{aligned}
\bq=\left\{{R}_{O_{B_X}},~{R}_{O_{B_Y}},~{R}_{O_{B_Z}},~\Phi,~\Theta,~\Psi;~\eta_{E_i}\right\}^{\rT}
\end{aligned}
\label{eq:generalized_coordinate}
\end{equation}
%
%
\par
Overall, it is more convenient to derive the equations of motion as a function of velocities expressed in the body reference frame. 
To relate $\bV_{O_B} $ to the linear velocity of the body frame origin expressed in the body frame $\bV_{{O_B},B}=\left\{U,V,W\right\}^{\rT}$ it is:
%
\begin{equation}
\bV_{O_B} =
\begin{aligned}
\left\{\begin{matrix}\dot{R}_{O_{B_X}}\\ \dot{R}_{O_{B_Y}}\\ \dot{R}_{O_{B_Z}} \end{matrix}\right\} = \bT_{I-B} \left\{\begin{matrix}U\\V\\ W\end{matrix}\right\}
\end{aligned}
\label{eq:kinematic1}
\end{equation}
%
where $\mbbm{T}_{I-B}$ is the direction-cosine matrix expressing the orientation of frame $\Sigma_B$ with respect to frame $\Sigma_I$.
\par
Finally, the Euler angles rates are related to the components of the angular velocity of Frame $\Sigma_B$ with respect to the frame $\Sigma_I$, $\bom_{B,I}=\left\{P,Q,R\right\}^{\rT}$, expressed in the frame $\Sigma_B$, by the well-known Euler equations:
%
\begin{equation}
\left\{
\begin{aligned}
& \dot{\Phi} = P +Q\sin{\Phi}\tan{\Theta}+R\cos{\Phi}\tan{\Theta}\\
& \dot{\Theta} = Q\cos{\Phi}-R\sin{\Phi}\\
& \dot{\Psi} = \left(Q\sin{\Phi}+R\cos{\Phi}\right)\sec{\Theta}
\end{aligned}
\right.
\label{eq:euler}
\end{equation}
%
%
\subsubsection{The Mean-Axes frame}   %
%
The equations governing the dynamics of the flexible aircraft can be remarkably simplified with an ad-hoc selection of the vehicle-fixed frame.
In references~\cite{Milne1962,Milne1968} it is noted that, for an elastic body, a coordinate frame always exists such that the linear and angular momenta relative to elastic deformations are identically zero.
This noninertial reference system is called \emph{mean-axes} frame and its origin coincides with the instantaneous CoM of the aircraft.
These axes move in phase with the body motion but they are not attached to any material point of the aircraft.
\par
By definition, the mean-axes constraints (or properties) are:
%
\begin{equation}
\left\{
\begin{aligned}
&\int_{\text{Vol}} \left. \frac{\ud \br}{\ud t} \right|_B \rho_\text{V}\ud\text{V} =  \bzero \\
&\int_{\text{Vol}} \br \x \left. \frac{\ud \br}{\ud t} \right|_B \rho_\text{V}\ud\text{V} = \bzero
\end{aligned}
\right.
\label{eq:mean_axes}
\end{equation}
%
where 
$\rho_\text{V}$ is the density of the material, \emph{Vol} represents the aircraft body, and $\left. \frac{\ud }{\ud t} \right|_B$ is the derivative as measured in the body frame. 
In reference~\cite{Canavin1977} it is shown how this is the body reference frame providing the minimum kinetic energy. 
%
\subsubsection{Final equations of motion} %
%
Considering the mean-axes frame constraints, as well as the hypothesis of \emph{linearly} behaving structure (i.e., small elastic deformations and deformations rate) and constant inertia tensor (despite the elastic deformations), expressions of the kinetic and potential energy strongly simplify (the interested reader is referred to works~\cite{Auricchio_MS,Castellanos2019MScThesis} for the derivation). 
Deriving with respect to the generalized coordinates, the equations of motion are obtained.
\paragraph{Rigid-body translation}
Deriving with respect to the  generalized coordinates relative to the vehicle CoM it is:
%
\begin{equation}
\begin{aligned}
&m \left(  \left.\frac{\ud \bV_{O_B}}{\ud t}\right|_B+\bom_{B,I} \x \bV_{O_B}\right)
=m \bg + \mathbf{Q}_{trasl}\\
\end{aligned}
\label{eq:lagrange_translation3}
\end{equation}
%
where $Q_{trasl}$ are the GAF relative to rigid translation of the vehicle. It is worth to underline that, due to the given assumptions,  Eq.~(\ref{eq:lagrange_translation3}) is formally identical to the one typically expected for a rigid body; 
effects of the structural deformations are only retained in the aerodynamic forces on the right-hand side. Hence, the aerodynamic operator is responsible for including effects of deformations. 
\paragraph{Rigid-body rotation}
Deriving with respect to the  generalized coordinates relative to the orientation of the body reference frame it is:
%
\begin{equation}
\begin{aligned}
\bI_0 \, \dot{\bom}_{B,I}+{\bom}_{B,I}\x \left( \bI_0 {\bom}_{B,I}\right) = \mathbf{Q}_{rot}
\end{aligned}
\label{eq:eqom_rotation_vectorial}   
\end{equation}
%
being $Q_{rot}$ are the GAF relative to rigid rotation of the vehicle and $\bI_0$ the tensor of inertia of the undeformed configuration.
To obtain Eq.~(\ref{eq:eqom_rotation_vectorial}),  Eq.~(\ref{eq:euler}) has been used to relate Euler's angles to the components of the angular velocity $\bom_{B,I}$.
Same as for the translational dynamics, effects of structural deformation are introduced into this equation by  aerodynamic forces $\mathbf{Q}_{rot}$. 
%
%
%
\paragraph{Elastic deformation}
Deriving with respect to the generalized coordinates relative to the elastic  modes $\boldsymbol{\eta}_{E}$ it is:
\begin{equation}
\begin{aligned}
\mathbf{M_{EE}} \; \ddot{\boldsymbol{\eta}}_{E} +\mathbf{C_{EE}} \; \dot{\boldsymbol{\eta}}_{E} + \mathbf{K_{EE}} \; \boldsymbol{\eta}_{E} = {\mathbf{Q}_{el}}
\end{aligned}
\label{eq:eom_deformation} 
\end{equation}
%
where $\mathbf{M_{EE}}$, $\mathbf{C_{EE}}$ and $\mathbf{K_{EE}}$ are the aircraft generalized mass, damping and stiffness matrices, and  $\mathbf{Q}_{el}$ are the GAF relative to the elastic modes.  
%
%
%
\subsubsection{Perturbation equations and stability analysis}       %
\label{sss:perturb1}                                                                   %
%
%
Small perturbation theory is applied to the equations of motion, in order to study stability properties of the system. 
%
Variables are decomposed into their reference and perturbation values, 
higher order terms are discarded and a set of linear equations is finally obtained.
%
%
%
\par
It is well known that the perturbation equations depend on the particular reference condition chosen to linearize about~\cite{Etkin_2012}. 
For this investigation, the reference configuration body axes coincide with the stability axes and only straight-and-level flight condition is considered, with zero sideslip angle:
all reference  velocities, Euler's angle and rotations are zero but the reference asymptotic airspeed in $x$ direction $V_\infty$.
%
%
In such conditions, the perturbation dynamics in the longitudinal and lateral-directional planes are decoupled, thus two independent sets of equations can be formulated and solved.
%
%
%

\subsection{Aerodynamic Model}     %
\label{ss:Aerodynamic}
%
%
%
%
\subsubsection{Enhanced Doublet Lattice Method}
\label{s:enhancedDLM}
%
The DLM~\cite{Rodden_1968,Rodden_Taylor_1998} is an assessed and still widely used method in aeroelasticity to evaluate the unsteady aerodynamic forces on deformable bodies. 
%
%
It is known, though, that in its original formulation, the DLM only calculates the unsteady aerodynamic loads and relative generalized forces due to local pitching and plunging 
of the aerodynamic surfaces, ignoring in plane forces and contributions due to  
in-plane motion. 
%
Moreover, the classic formulation does not take into account contributions to perturbation forces due to nonzero loads in the reference condition (i.e., loads distribution in trim).
%
%
%
While the approximation delivers good results for most of applications on conventional wings, it fails to provide reliable loads evaluation on some configurations, as for example, T-tails~\cite{VanZyl-2011}: 
%
in this case, both in-plane loads and normal loads due to in-plane motions are important to be modeled. 
It is clear that, for the PrandtlPlane, due to its particular layout, inclusion of these terms may have a relevant impact, especially for lateral-directional stability analyses.
%
%
\par
Enhancements to the traditional DLM are implemented in the in-house solver used within this investigation, following efforts~\cite{VanZyl-2011,VanZyl-2008,Kier2011AnIL,Kier2009UNIFYINGMA}. 
%
%
A more general form of the aerodynamic boundary condition can be written, for harmonic motion, as:
\begin{equation}
\begin{aligned}
\mathbf{u}_{1} \cdot \mathbf{n}_0 = i \omega \;\mathbf{h} \cdot \mathbf{n}_0  - \mathbf{u}_{\infty} \cdot (\mathbf{r} \times \mathbf{n}_0)
\end{aligned}
\label{eq:boundary_perturb} 
\end{equation}
%
where $\mathbf{h}$ and $\mathbf{r}$ are, respectively, the panel displacement and rotation vectors for the considered motion at frequency $\omega$; $\mathbf{n}_0$ is the panel mean normal vector; $\mathbf{u}_{\infty}$ is the free stream velocity vector and $\mathbf{u}_{1}$ is the unsteady perturbation velocity vector, as seen by the aerodynamic panel~\cite{VanZyl-2011}.
%
In the original DLM formulation $\mathbf{u}_{\infty}$ is always considered directed along the panel chord, whereas now the reference condition of the aircraft is taken into account and $\mathbf{u}_{\infty}$ and $\mathbf{n}_0$ are, in general, not perpendicular. 
Hence, the second term on the right hand side of Eq.~(\ref{eq:boundary_perturb}) features 
%
an extra contribution proportional to the angle of the attack of the wing and which takes into account yaw/dihedral coupling.
%
%
%
\par
%
Enhancements concerning forces evaluation can be better visualized using the Kutta-Joukowsky law in it vector form. 
%
The \textit{first harmonic} of the perturbation forces on a lifting surface panel is given by~\cite{VanZyl-2011}:
%
%
\begin{equation}
\begin{aligned}
\mathbf{F_1} =  \rho \, \left[ \mathbf{u}_{\infty} \times \mathbf{\Gamma}_1 + \mathbf{u}_{\infty} \times (\mathbf{r} \times \mathbf{\Gamma}_0) + \omega \mathbf{h} \times  \mathbf{\Gamma}_0 \right]
\end{aligned}
\label{eq:Kutta_unsteady} 
\end{equation}
%
where $\Gamma_0$ is the steady circulation in the reference condition while $\Gamma_1$ is the unsteady circulation due to the harmonic motion. It can be noted that the traditional DLM only takes into account the out-of-plane component of the first term on the right hand side.
%
%
%
In the current extended formulation, the first term now properly predicts the out-of-plane component and the relative in-plane forces due to the local angle of attack in the reference condition.
%
With respect to the second and third terms on the right hand side of  Eq.~(\ref{eq:Kutta_unsteady}), they depend on the circulation vortices in the reference condition $\Gamma_0$, which, in turn, depend on the forces needed to trim the aircraft in the sought flight condition. 
With these terms it is possible to model aerodynamic forces which are relevant, among others, for yaw-roll coupling effects. However, it is worth mentioning that the  second term in the right hand side of Eq.~(\ref{eq:Kutta_unsteady}) has not been included, as it has been shown that its integration in the Generalized Force calculation procedure, in isolation from the quadratic mode shape contributions~\cite{VanZyl-2012}, may lead to physically wrong results.
%
%
\par
Once the reference flight condition is selected, calculation of $\Gamma_0$ and the angle of attack is carried out by the VLM used to evaluate zero-frequency contributions to the unsteady perturbation forces~\cite{Rodden_1968}. This allows to correctly integrate the so-far discussed contributions in the aerodynamic boundary condition and in the forces term.
%
%
%
%
%
\paragraph{Generalized Aerodynamic Forces}
%
The GAF  $\tilde{\mathbf{Q}}$ 
acting on a thin lifting system are given for a harmonic oscillation of the structure at a certain reduced frequency $k$:
%
\begin{equation}
\begin{aligned}
\widetilde{\bf{Q}} = q_\infty \widetilde{\bA}\left({ik}\right)  \widetilde{\boldsymbol{\eta}}
\end{aligned}
\label{eq:RFA}
\end{equation}
%
where $\widetilde{\boldsymbol{\eta}}$ is the generalized modal coordinate vector, which includes both rigid ($\widetilde{\boldsymbol{\eta}}_R$) and elastic ($\widetilde{\boldsymbol{\eta}}_E$) terms, $q_\infty$ is the dynamic pressure and $\widetilde{\bA}\left({ik}\right)$ is the GAF matrix. 
The tilde is here used to specify that these generalized forces are relative to the generalized coordinates of the normal modes-based decomposition performed with respect to a reference system (i.e., the DLM one) which is, in general, different than the stability axes used in Eqs.~(\ref{eq:lagrange_translation3}) to (\ref{eq:eom_deformation}). A transformation is needed to make them consistent with the primary Lagrangian coordinates $\bq$, as it will be discussed in section~\ref{sss:transf1}.
The  reduced frequency $k$ is obtained by normalizing the circular frequency $\omega$ with the free-stream velocity $V_\infty$ and the reference length $\hat{c}$, i.e., $k=\frac{\omega \, V_\infty }{ \hat{c} }$).
%
%
\subsubsection{Rational Function approximation} \label{sss:rational}
The linear unsteady aerodynamic operator is of the transcendental kind in the Laplace domain, due to time delays in the propagation of disturbances (compressibility effects) and the convected wake vorticity~\cite{Gennaretti_MastroddiJoA2004}.
To overcome this difficulty, the operator is approximated by rational expressions involving a finite number of poles; in the state-space representation, thus, aerodynamic forces are described by a finite number of states.
Typically, it holds:
%
\begin{equation}
\begin{aligned}
\widetilde{\bA} (i k)  = \widetilde{\bA}_0  +
ik \, \widetilde{\bA}_1 +\left(i k  \right)^2 \widetilde{\bA}_2  +\text{High Order Terms}
\end{aligned}
\label{eq:RFA2}
\end{equation}
%
The part without Higher Order Terms (HOT) represents a \emph{quasi-steady} approximation. Popular procedures to determine the HOT in Eq.~(\ref{eq:RFA2}) are the so-called Rational Function Approximation (RFA) \cite{Roger1977,Karpel1982}, or alternatively, Rational Matrix Approximation (RMA) \cite{Morino_Mantegazza_1995AIAAJ} methods.
The approach considered in this paper is based on Roger's work~\cite{Roger1977}, for which:
\begin{equation}
\begin{aligned}
\widetilde{\bA} (i k)  = \widetilde{\bA}_0  +
ik \, \widetilde{\bA}_1 +\left(i k  \right)^2 \widetilde{\bA}_2  + \sum_{j=1}^{N_{\text{lag}}} \frac{ik}{ik+ \frac{\beta_j \hat{c}}{V_\infty}  } \widetilde{\bA}_{2+j}
\label{eq:Roger}
\end{aligned}
\end{equation}
%
where $\beta_i$ are lag coefficients set by the user.
\par 
The DLM calculates $\widetilde{\bA}$ for a selected set of reduced frequencies $k$ (for a given Mach number). 
To fit the provided data and  express $\widetilde{\bA}$ as shown in Eq.~(\ref{eq:Roger}), a least-square approach is pursued.  
Typically, the steady values $\widetilde{\bA}(0)$  determine  $\widetilde{\bA}_0$, and the fitting procedure involves matrices  $\widetilde{\bA}_j$ with $j>0$. 
\par 
It is important to note that the fitting procedure can be fine-tuned to better approximate some regions of frequencies. 
To increase precision in the $k\rightarrow0$ region a modified Roger method
is here implemented, which adds to the original set of Least-Square constraints
%
a further condition on the \emph{derivative} of the interpolated GAF matrix, as better explained in~\cite{UC3Mteam_PrP1ceas}.
%
This method has proven to be useful to increase accuracy around the $k \rightarrow 0$ region of frequencies, where most of the flight-dynamic physics is held, preserving, at the same time, good interpolation for higher frequencies characteristic of the aeroelastic behaviour. 
%
%
\par 
%
Neglecting higher-order terms, it holds that $\widetilde{\bA}_1 = \left. \frac{\ddi {\widetilde{\bA}}}{\ddi \,  ik} \right|_{k=0}$, and $\widetilde{\bA}_2 = \frac{1}{2} \left. \frac{\ddi^2 {\widetilde{\bA}}}{\ddi \,  (ik)^2} \right|_{k=0}$. 
Notice how this justifies the terminology \emph{quasi-steady}: the model is adequate in the region of small reduced frequencies. 
%
\subsubsection{Analytical continuation}
Invoking the \emph{analytical continuation}~\cite{Dettman_BOOK1969}, if an analytical function is known in terms of the imaginary variable $ik$, the same expression with $ik$ can be used for a the generic complex variable $p = g + ik$, and the function is finally described in the whole complex plane.
Thus, the above expressions can be extended just substituting $p$ (nondimensional Laplace variable) to $ik$.
%
%
\subsubsection{Rigid generalized coordinates and rigid modes}
\label{sss:transf1}
%
In order to be consistent with Eq.~(\ref{eq:lagrange}), 
the modal rigid-body coordinates of Eq.~(\ref{eq:RFA}) used in the DLM, needs to be expressed in terms of the rigid Lagrangian coordinates of Eq.~(\ref{eq:generalized_coordinate}). 
%
%
%
Such transformation depends on the reference flight condition; for the chosen one  (straight-and-level flight in stability  axes) and for the the longitudinal plane dynamics, it holds: 
%
\begin{equation}
\begin{aligned}
\widetilde{\boldsymbol{\eta}}_R =
\left\{{\begin{array}{c}
	\widetilde{\eta}_{R_x}\\
	\widetilde{\eta}_{R_z}\\
	\widetilde{\eta}_{R_{\theta_y}}\\
	\end{array}}\right\} = \underbrace{\left[\begin{array}{ccc}
\cos(\pi - AoA_{}) & \sin(\pi - AoA_{}) & 0  \\
-\sin(\pi - AoA_{}) & \cos(\pi - AoA_{}) & 0  \\
0 & 0 & 1 \\
\end{array}\right]}_{\bT_1}
\left\{{\begin{array}{c}
	x_E\\
	z_E\\
	\theta\\
	\end{array}}\right\}  
\end{aligned}
\label{eq:rmodes2rigidvector1_long}
\end{equation}
%
%
\begin{equation}
\begin{aligned}
\boldsymbol{\dot{\widetilde{\eta}}}_R =
\left\{{\begin{array}{c}
   \dot{\widetilde{\eta}}_{R_x}\\
   \dot{\widetilde{\eta}}_{R_z}\\
   \dot{\widetilde{\eta}}_{R_{\theta_y}}\\
\end{array}}\right\} =& 
\underbrace{\left[\begin{array}{ccc}
0 & 0 & 0  \\
0 & 0 & V_\infty  \\
0 & 0 & 0 \\
\end{array}\right]}_{\bT_2} 
 \left\{{\begin{array}{c}
   x_E\\
   z_E\\
   \theta\\
\end{array}}\right\}+
\bT_3
 \left\{{\begin{array}{c}
   u\\
   w\\
   q\\
\end{array}}\right\}
\end{aligned}
\label{eq:rmodes2rigidvector2_long}
\end{equation}
%
%
\begin{equation}
\begin{aligned}
\boldsymbol{\ddot{\widetilde{\eta}}}_{R} =
\left\{{\begin{array}{c}
	\ddot{\widetilde{\eta}}_{R_x}\\
	\ddot{\widetilde{\eta}}_{R_z}\\
	\ddot{\widetilde{\eta}}_{R_{\theta_y}}\\
	\end{array}}\right\} =& \underbrace{\left[\begin{array}{ccc}
0 & 0 & 0  \\
0 & 0 & V_\infty  \\
0 & 0 & 0 \\
\end{array}\right]}_{\bT_4}
\left\{{\begin{array}{c}
	\dot{x}_E\\
	\dot{z}_E\\
	\dot{\theta}\\
	\end{array}}\right\}+
\bT_5
\left\{{\begin{array}{c}
	\dot{u}\\
	\dot{w}\\
	\dot{q}\\
	\end{array}}\right\}
\end{aligned}
\label{eq:rmodes2rigidvector3_long}
\end{equation}
%
%
%
where $\bT_5 =\bT_3 = \bT_1$,  $\widetilde{\eta}_{R_x}$, $\widetilde{\eta}_{R_z}$ and $\widetilde{\eta}_{R_{\theta_y}}$ are the modal coordinates of the three rigid modes in the longitudinal plane, and $AoA$ is the angle of attack, in radiant, for given asymptotic speed and mass of the vehicle with respect to the DLM reference system. 
It is worth underlying that the DLM reference system features the $x$-axis directed along the unperturbed asymptotic speed and $y$-axis along the span-wise direction. 
%
%
Recall that  Eqs.~(\ref{eq:rmodes2rigidvector1_long}) to (\ref{eq:rmodes2rigidvector3_long}) 
hold if the rigid-body modes are selected to be of unitary translation/rotation, otherwise a scaling factor needs to be applied to the matrices to recover consistency with the generalized coordinates of Eq.~(\ref{eq:generalized_coordinate}).
The reader is referred to work~\cite{UC3M_UFFDScitech} for the expression relative to the lateral-directional dynamics.
%
%
%
\subsubsection{Expressing the GAFs in the stability axes}
%
The generalized forces obtained with the DLM need to be projected to be consistent with the Lagrangian coordinates of Eq.~(\ref{eq:generalized_coordinate}). 
With reference to Eq.~(\ref{eq:lagrange}) it holds:
\begin{equation}
{\bf{Q}} =  \frac{\pd  \mathcal{W}}{\pd \bq}^T = \left[ \frac{\pd  \mathcal{W}}{\pd \boldsymbol{\widetilde{\eta}} } \cdot \frac{\pd \boldsymbol{\widetilde{\eta}} } {\pd \bq} \right]^T = \frac{\pd \boldsymbol{\widetilde{\eta}} } {\pd \bq}^T \cdot \frac{\pd  \mathcal{W}}{\pd \boldsymbol{\widetilde{\eta}} }^T
=\mathcalbf{R}^T\widetilde{\bf{Q}}= q_\infty \underbrace{ \mathcalbf{R}^T \widetilde{\bA} \left({ik}\right)}_{\bA(ik)} \boldsymbol{\widetilde{\eta}}
\label{eq:rotated_forces}
\end{equation}
%
%
For the longitudinal dynamics $\mathcalbf{R}$ is:  
\begin{equation}
\mathcalbf{R} = \left[ \begin{array}{cc}
\bT_1
& \boldsymbol{0}^{3 \x n_E}  \\ 
\boldsymbol{0}^{n_E \x 3}  & \boldsymbol{I}^{n_E} 
\end{array}\right]
\label{eq:rotation_matrix_long}
\end{equation}
\par 
In the case of the quasi-steady approximation, applying the inverse Fourier transform, the GAFs can be expressed as: 
\begin{equation}
{\bf{Q}} =  q_\infty \left( {\bA}_0  \, \boldsymbol{\widetilde{\eta}}  + \frac{\hat{c}}{V_\infty }{\bA}_1  \, \boldsymbol{\dot{\widetilde{\eta}}}  +  \frac{\hat{c}^2}{V_\infty^2}  {\bA}_2 \, \boldsymbol{\ddot{\widetilde{\eta}}}   \right)
\label{eq:faero_FT}
\end{equation}
%
%
%

\subsection{State-Space Formulation} %
%
In this section both the sets of equations governing the dynamics of the rigid and flexible aircraft are considered. 
Considering that the coupling between rigid and elastic dynamics is introduced by the aerodynamic forces only, to model the dynamics of the flexible aircraft it is formally sufficient to augment system of Eqs.~(\ref{eq:lagrange_translation3}) and (\ref{eq:eqom_rotation_vectorial}) with Eq.~(\ref{eq:eom_deformation}) and extending the set of Lagrangian coordinates. 
%
Only specialization for the longitudinal case is shown, reader is referred to~\cite{UC3M_UFFDScitech} for the expression relative to the lateral-directional dynamics.
%
\subsubsection{Rigid part of the aircraft dynamics}
\label{sss:rigidpart}
%
%
%
The set of perturbation equations in state-space form governing the rigid part of the aircraft dynamics in stability axes~\cite{Etkin_2012} is:
%
\begin{equation}
\begin{aligned}
& \left[\begin{array}{cccccc}
m & 0 & 0 & 0 & 0 & 0 \\
0 & m & 0 & 0 & 0 & 0 \\
0 & 0 & I_{yy} & 0 & 0 & 0 \\
0 & 0 & 0 & 1 & 0 & 0 \\
0 & 0 & 0 & 0 & 1 & 0 \\
0 & 0 & 0 & 0 & 0 & 1 \\
\end{array}
\right] \left\{{\begin{array}{c}
	\dot{u}\\
	\dot{w}\\
	\dot{q}\\
	\dot{x}_E\\
	\dot{z}_E\\
	\dot{\theta}\\
	\end{array}}\right\} =
& \left[\begin{array}{cccccc}
0 &  0 & 0  & 0 & 0 & -mg   \\
0 & 0 & mV_\infty & 0 & 0 & 0 \\
0 & 0 & 0 & 0 & 0 & 0 \\
1 & 0 & 0 & 0 & 0 & 0 \\
0 & 1 & 0 & 0 &  0 & -V_\infty \\
0 & 0 & 1 & 0 & 0 & 0 \\
\end{array}
\right]
\left\{{\begin{array}{c}
	u\\
	w\\
	q\\
	x_E\\
	z_E\\
	\theta\\
	\end{array}}\right\}+\left\{{\begin{array}{c}
	{Q}_x \\
	{Q}_z \\
	{Q}_{\theta_y} \\
	0 \\
	0 \\
	0 \\
	\end{array}}\right\}
\end{aligned}
\label{eq:salf_matrix_long2}
\end{equation}
%
being the considered rigid state:
%
\begin{equation}
\begin{aligned}
\bx_R = \left\{{\begin{array}{c}
	u\\
	w\\
	q\\
	x_E\\
	z_E\\
	\theta\\
	\end{array}}\right\} 
\hspace{.8cm}
\begin{array}{l}
\rightarrow \text{longitudinal~velocity}   \\ 
\rightarrow \text{vertical~velocity}\\
\rightarrow \text{pitch~rate} \\
\rightarrow \text{longitudinal~position} \\
\rightarrow \text{vertical~position} \\
\rightarrow \text{2nd~Euler~angle}
\\
\end{array}
\end{aligned}
\label{eq:rigid_state_salf_long}
\end{equation}
%
%
%
\par
The above system of Eq.~(\ref{eq:salf_matrix_long2}) can be expressed as:
\begin{equation}
\begin{aligned}
\bM_{RR} \;  \dot{\bx}_R = \bK_{RR} \; \bx_R + \mathbf{Q}_{R}
\end{aligned}
\label{eq:rigid_matrix_salf2_long}
\end{equation}
%
%
%
\subsubsection{Rigid aircraft with quasi-steady aerodynamics}
%
%
With respect to system of Eq.~(\ref{eq:rigid_matrix_salf2_long}), let's consider the quasi-steady form of the GAF expressed in terms of generalized coordinates of Eq.~(\ref{eq:faero_FT}). 
Being only the rigid modes involved, the GAF matrices $\bA_i$ are indicated with $\bA_{i_{RR}}$. 
The aerodynamic force vector becomes then:
\begin{equation}
{\mathbf{Q}}_R =  q_\infty \left( \bA_{0_{RR}}  \, \boldsymbol{\widetilde{\eta}}_R  + \frac{\hat{c}}{V_\infty } \bA_{1_{RR}} \, \boldsymbol{\dot{\widetilde{\eta}}}_R  +  \frac{\hat{c}^2}{V_\infty^2}  \bA_{2_{RR}} \, \boldsymbol{\ddot{\widetilde{\eta}}}_R   \right)
\label{eq:faero_FT2}
\end{equation}
%
%
Transforming the modal rigid-body coordinates (Eqs.~(\ref{eq:rmodes2rigidvector1_long}), (\ref{eq:rmodes2rigidvector2_long}) and (\ref{eq:rmodes2rigidvector3_long})) and recalling the expression of the state vector of Eq.~(\ref{eq:rigid_state_salf_long}) it is possible to obtain an expression suitable for the state-space form:
%
\begin{multline}
\label{eq:rigidquasisteady}
{\mathbf{Q}}_R =  q_\infty \left( \left[ 
\begin{array}{cc}
\boldsymbol{0}^{n_R \times 3}  &  \bA_{0_{RR}} \bT_1 
\end{array}
\right] +   
\frac{\hat{c}}{V_\infty}   \left[ 
\begin{array}{cc}
\bA_{1_{RR}} \bT_3  &   \bA_{1_{RR}} \bT_2 
\end{array}
 \right]   
\right)  \bx_R^{}  + \\
q_\infty \left(   
\frac{\hat{c}^2}{V^2_\infty}   \left[ 
\begin{array}{cc}
\bA_{2_{RR}} \bT_5  &   \bA_{2_{RR}} \bT_4 
\end{array}
\right]   
\right)  \dot{\bx}_R^{}
\end{multline}
%
The state-space system is then completely defined, and stability analysis can be trivially performed. 
%
%
\subsubsection{Flexible aircraft with quasi-steady aerodynamics}
\label{sss:flexible_aerodynamics}
%
To take into account the dynamics of the flexible aircraft, Eq.~(\ref{eq:eom_deformation}) is added to the system of Eq.~(\ref{eq:rigid_matrix_salf2_long}) and the rigid state vector $\bx_R$ is augmented with the elastic generalized coordinates $\boldsymbol{\eta}_{E}$, obtaining: 
%
\begin{equation}
\begin{aligned}
\bx^{} = \left\{{\bx_R^{T}}; \dot{\boldsymbol{\eta}}_{E}^T ,  \boldsymbol{\eta}_{E}^T \right\}^T
\end{aligned}
\label{eq:rigid_matrix_salf3_lateral}
\end{equation}
%
Consider now the quasi-steady form of the GAFs of Eq.~(\ref{eq:faero_FT}). 
It is convenient to partition the matrices $\bA_i$ in terms of GAFs relative to rigid ($R$) and elastic ($E$) modes. It holds, then:
\begin{equation}
\bA_i = \left[ \begin{array}{cc}
 \bA_{i_{RR}} &  \bA_{i_{RE}}  \\
 \bA_{i_{ER}} &  \bA_{i_{EE}}
\end{array}\right]
\label{eq:aero_coupling}
\end{equation}
%
Expressing the modal rigid-body coordinates in terms of rigid Lagrangian coordinates, Eqs.~(\ref{eq:rmodes2rigidvector1_long}), (\ref{eq:rmodes2rigidvector2_long}) and (\ref{eq:rmodes2rigidvector3_long}),  it is possible to refer the aerodynamic forces in terms of the state vector. %
The problem can be recast in the following state-space form:
%
%
\begin{equation}
\resizebox{\hsize}{!}{$
\begin{aligned}
\left[\begin{array}{c|c|c}
\left({\bM_{RR}^{}}-q_\infty\left(\frac{\hat{c}}{V_\infty}\right)^2\left[{\begin{array}{cc}
   \bA_{2_{RR}} \bT_5 & \bA_{2_{RR}} \bT_4\\
   \boldsymbol{0}^{n_R} & \boldsymbol{0}^{n_R} \end{array}}
   \right]\right) &
   \left[\begin{array}{c}
-q_\infty\left(\frac{\hat{c}}{V_\infty}\right)^2 \bA_{2_{RE}} \\
\boldsymbol{0}^{n_R \x n_E} 
\end{array} \right] &
\boldsymbol{0}^{2n_R \x n_E}  \\ \\
\hline
-q_\infty\left(\frac{\hat{c}}{V_\infty}\right)^2 \bA_{2_{ER}} \left(\bT_5~~\bT_4 \right) &
\left( \mathbf{M_{EE}} -q_\infty\left(\frac{\hat{c}}{V_\infty}\right)^2 \bA_{2_{EE}}\right)&
\boldsymbol{0}^{n_E} \\ \\ \hline
\boldsymbol{0}^{n_E \x 2n_R}  & \boldsymbol{0}^{n_E}  & \boldsymbol{I}^{n_E} 
\end{array} \right]
\left\{{\begin{array}{c}
   \dot{\bx}_R \\
   \ddot{\boldsymbol{\eta}}_E\\
   \dot{\boldsymbol{\eta}}_E\\
\end{array}}\right\}=
\\ \\
\left[\begin{array}{c|c|c}
 \bK_{RR}^{} + q_\infty\left(

   \left[{\begin{array}{cc}
   \frac{\hat{c}}{V_\infty}\bA_{1_{RR}} \bT_3 & \frac{\hat{c}}{V_\infty}\bA_{1_{RR}} \bT_2 + \bA_{0_{RR}} \bT_1 \\
   \boldsymbol{0}^{n_R} & \boldsymbol{0}^{n_R} \end{array}}
   \right]
   \right) &  \left[\begin{array}{c}
\frac{q_\infty \hat{c}}{V_\infty}\bA_{1_{RE}} \\
\boldsymbol{0}^{n_R \x n_E }
\end{array} \right] & \left[\begin{array}{c}
q_\infty \bA_{0_{RE}} \\
\boldsymbol{0}^{n_R \x n_E }
\end{array} \right]\\ \\ \hline
q_\infty\left(
   \left[{\begin{array}{cc}
   \frac{\hat{c}}{V_\infty}\bA_{1_{ER}} \bT_3 & \frac{\hat{c}}{V_\infty}\bA_{1_{ER}} \bT_2 +\bA_{0_{ER}} \bT_1 \end{array}}
   \right]
   \right) &
   -\left( \mathbf{C_{EE}} -\frac{q_\infty \hat{c}}{V_\infty} \bA_{1_{EE}} \right) &
   -\left( \mathbf{K_{EE}} -q_\infty \bA_{0_{EE}} \right) \\ \\ \hline
   \boldsymbol{0}^{n_E \x 2n_R }  &  \boldsymbol{I}^{n_E}  & \boldsymbol{I}^{n_E}
\end{array} \right]
\left\{{\begin{array}{c}
   {\bx}_R \\
   \dot{\boldsymbol{\eta}}_E\\
   {\boldsymbol{\eta}}_E\\
\end{array}}\right\}
\end{aligned} $
}
\label{eq:matrix_nolag}
\end{equation} 
%

The same approach can be used for the lateral-directional dynamics as detailed in efforts~\cite{UC3M_UFFDScitech,Castellanos2019MScThesis}.
\par
When unsteady aerodynamics is considered, the system is augmented with aerodynamic added states.  
The final state-space system  depends on the chosen model-lag augmentation technique (RFA/RMA), and, for the sake of brevity, it is here omitted.  
The interested reader is referred to effort~\cite{Auricchio_MS} for more details.
%

%
\section{Computational Tools}      %
\label{s:computationaltool}
%
\subsection{UFFD}    %
%
%
Unified Flexible Flight Dynamics (UFFD) is an in-house tool implementing the formulation proposed in section~\ref{s:FDDA}. 
It builds on an in-house DLM developed by the authors and collaborators~\cite{Demasi_Livne_2009,SDSUteam_5jour} and whose capabilities have been extensively validated in the last decade against the commercial code NASTRAN~\cite{NASTRAN_aeroelastic}. The DLM tool closely follows the theoretical approach shown in  works~\cite{Rodden_1968,Rodden_Taylor_1998}. 
For the purpose of this investigation, the DLM has been enhanced to model the terms shown and discussed in section~\ref{s:enhancedDLM}.
%
%
%
%
\par 
Within the framework, to interpolate the GAF coefficients both Roger's \cite{Roger1977} and Karpel's \cite{Karpel_JoA1996} methods are available.  In this paper, though, only Roger method has been considered. 
\par
The finite element in-house code needed to perform the modal analysis is part of a more complex software capable of modeling geometric nonlinearities and perform post-critical analyses. The structural module is partially based on the work presented in~\cite{LevyBook}. 
\par 
To transfer aerodynamic forces/structural displacements to the structural/aerodynamic grid, an in-house interface toolbox has been used. 
This framework currently offers two types of projection methods, one based on Infinite Plate Splines (IPS), as described in reference~\cite{Harder_Desmarais_1972}, and another based on Moving Least Square (MLS) approach, see efforts~\cite{Mantegazza2,SDSUteam_5jour,Bombardieri_uc3m_EUROGEN}. 
In this investigation the IPS method is used. 
With respect to the traditional DLM method, which only requires pitch and plunge motions of the aerodynamic panels and only evaluates out-of-plane forces, the spline method has been extended to handle all extra motion components required by the enhanced DLM (i.e., in-plane displacements and roll- and yaw-like rotations of the aerodynamic panel). 
%
%
\par 
UFFD allows for great flexibility, performing several types of analyses. It is straightforward to switch between analyses for  rigid or flexible aircraft, or to select increasing level of aerodynamic complexities (from quasi-steady approach to fine-tuned RFAs or from the traditional DLM to the enhanced one).    
%
%

%
\section{Test Case: 250-seat PrandtlPlane}      %
\subsection{Description of the baseline configuration} %
\begin{figure}[htb]
     \centering
     \vspace{5mm}
         \includegraphics[width=\textwidth]{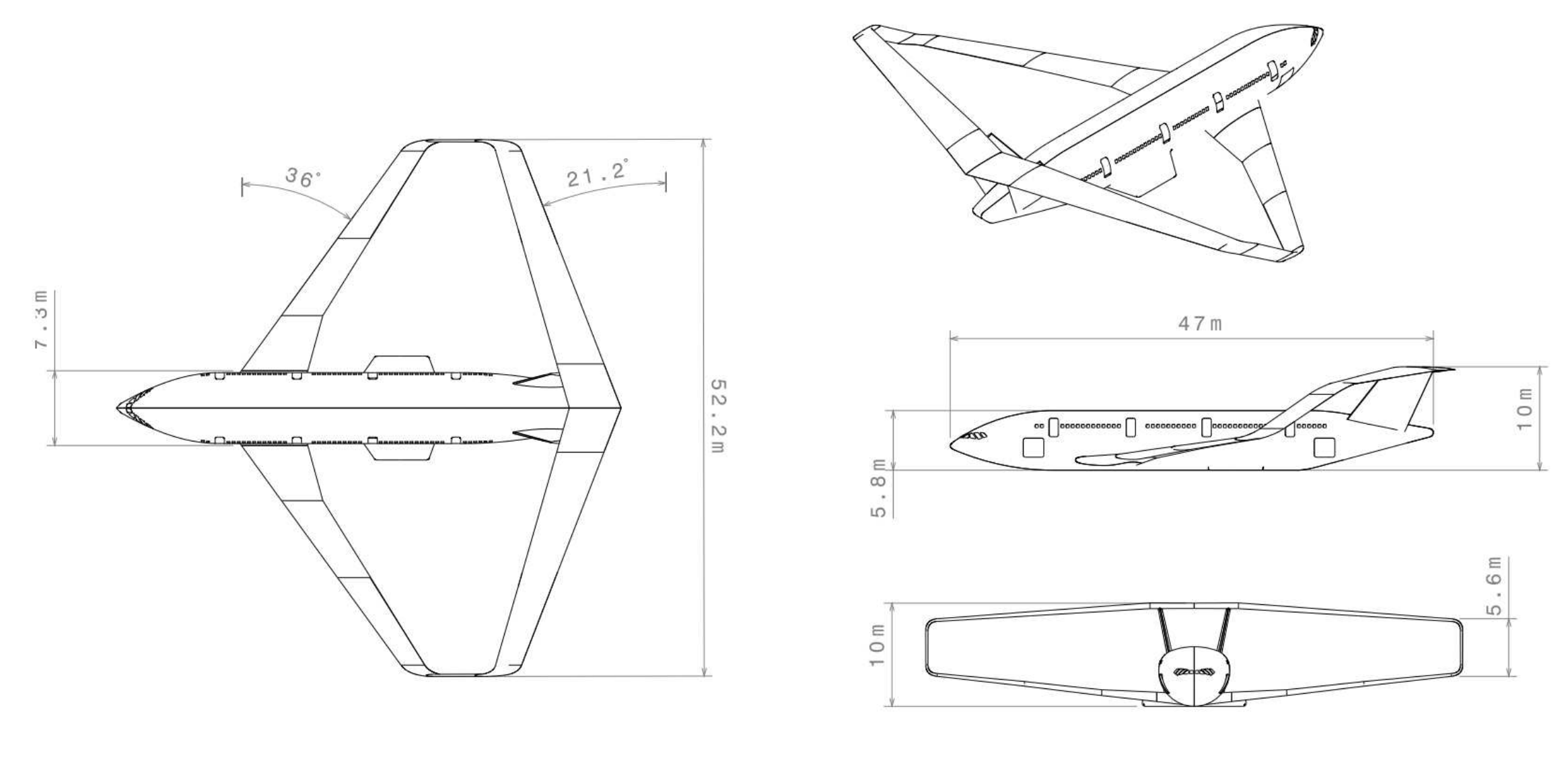}
         \caption{PrP250 concept, studied at University of Pisa~\cite{Frediani_Rizzo_etal_PrP250_AIDAA2005,Bottoni_Scanu_MSc_2004}.}
         \label{f:PrP250}
\end{figure}

%
\label{s:testacase}
The baseline of this study is the PrandtlPlane designed in reference~\cite{Frediani_Structural_Alluminium_SpringerII}. 
The vehicle, a 250-passenger concept, was first developed in effort~\cite{Frediani_Rizzo_etal_PrP250_AIDAA2005} to demonstrate the feasibility of the project: design mission data are given in Tab.~\ref{t:design_mission_data}
%
while an artistic view can be appreciated in Fig.~\ref{f:PrP250}; for the rest of the work it will be referred to as \emph{PrP250}.
%
\begin{table}[!t]
\label{t:design_mission_data}       
\centering
\begin{tabular}{ l l  }
\hline\noalign{\smallskip}
Range & 6000 nm (11112 km) \\
Take-off field length & 3000 m \\
Take-off airport altitude & sea level \\
Landing field length & 3000 m \\
Approach speed & 140 kts (260 km/h)\\ 
Cruise altitude & 10500 m\\
$(T/W)_{TO}$ & 0.254 \\
$(W/S)_{TO}$ & 0.575 kg/m$^2$ \\
$C_{D_O}$ & 0.0255 \\
$S_{REF}$ & 362.6 m$^2$ \\
\noalign{\smallskip}\hline
\end{tabular}
\caption{PrP250 design mission, from~\cite{Frediani_Rizzo_etal_PrP250_AIDAA2005}.}
\end{table}

%
\par 
The external surface and general layout shape were obtained by means of a multidisciplinary design optimization approach (see references~\cite{Rizzo_book,Rizzo_Springer} for a description of the framework).
Later, the structural design was fine-tuned~\cite{Frediani_Structural_Alluminium_SpringerII} taking into account different constraints as maximum stress, local buckling of stiffened panels, aileron efficiency, static aeroelasticity and flutter.
Further refinements and detailed design of the wing-box were carried out later~\cite{Frediani_composite_springerII}.
Among others, a solution adopting composite materials was proposed.
\par
Control surfaces were sized in references~\cite{Ginneken_AIAA2010-3060,Voskuijl_EriceII}; the design was carried out complying with a set of handling qualities requirements at several flight conditions (low speed, high speed, crosswind, high altitude, low altitude).
The handling qualities requirements took into account maneuvers as push-pull, minimum time to bank, aircraft trim with one engine out, take-off rotation and steady turn.
\par
The PrP250 concept, with respect to the configuration studied in the PARSIFAL project, has a larger range for a comparable payload, and presents a larger span of the wing system. 
Differently than what stated in~\cite{Picch-2020} it cannot be considered an obsolete design.
Moreover, due to its combination of stiffness and inertia distributions, aeroelastic phenomena on the PrP250 are relevant, and are, thus, more challenging to be studied and understood.
%
\subsubsection{Previous aeroelastic studies on the PrP250} %
%
This configuration has been the object of few preliminary aeroelastic studies. 
Gust response of the PrP250 and relative effects on the structural sizing, in terms of weight, were preliminary assessed in effort~\cite{UC3Mteam_PrP2ceas}.
Dynamic aeroelastic instabilities have been studied in references~\cite{Frediani_Flutter_SpringerII,SDSUteam_6jour}, for the longitudinal response, and in reference \cite{SDSUteam_8scitech} for the lateral-directional one. 
Anyway, the studies didn't feature the unified formulation discussed in this work, and the rigid modes didn't have the proper physical meaning of flight-dynamic modes.
In conference efforts~\cite{UC3Mteam_PrP1ceas,UC3M_UFFDScitech} a preliminary introduction of flight-dynamic aspects into the response of the free-flying and flexible aircraft was carried out.
%
%
%
%
%

%
%
%
%
%
%
\subsection{Aeroelastic model}                         %
\begin{figure}[htb]
	\centering
	\vspace{5mm}
	\includegraphics[width=\textwidth]{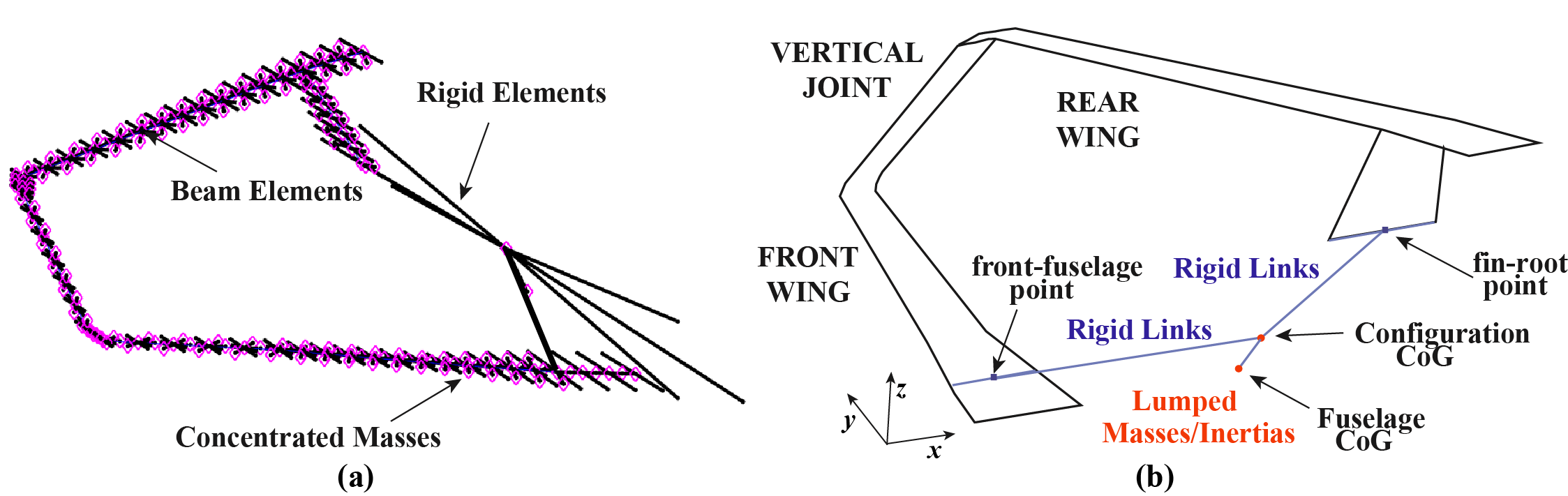}
	\caption{PrP250 computational model: (a) structural model and (b) wing-configuration sketch.}
	\label{f:PrP_fem}
\end{figure}

%
The computational model is illustrated in this section.
The structural model is described by finite elements
whereas the aerodynamic one by aerodynamic lifting surface panels.
A view of the finite element model is given in Fig.~\ref{f:PrP_fem}(a). 
The wing structure is described by beam elements with properties describing stiffness and inertia of the wing-box.
Concentrated inertial elements are also included to model structural (e.g., ribs), fuel and non-structural system inertia. 
multi-point-constraint equations are employed to connect concentrated inertial elements (slaves) with the wing-box structural nodes (masters). 
These equations enforce a displacement on the slave nodes which is the weighted average of the displacements of the masters. 
Rigid elements are used to model fuselage, the wing system-fuselage connections and also to provide a support to the splining process between the topologically different structural and aerodynamic meshes. 
%
%
\par
%
Concentrated nodal masses and inertias are also placed to reproduce fuselage structural and nonstructural weights, including payload; landing gear contribution is directly applied to the configuration CoM; engines inertias are located in the rear fin area. 
A system of rigid links connects these concentrated inertias to the front wing and fins, as clarified in  Fig.~\ref{f:PrP_fem}(b). 
Fuselage inertial properties have been extrapolated from data reported in work~\cite{Ginneken_MS2009}.
%
%
\par
\begin{figure}[htb]
	\centering
	\vspace{5mm}
	\includegraphics[width=\textwidth]{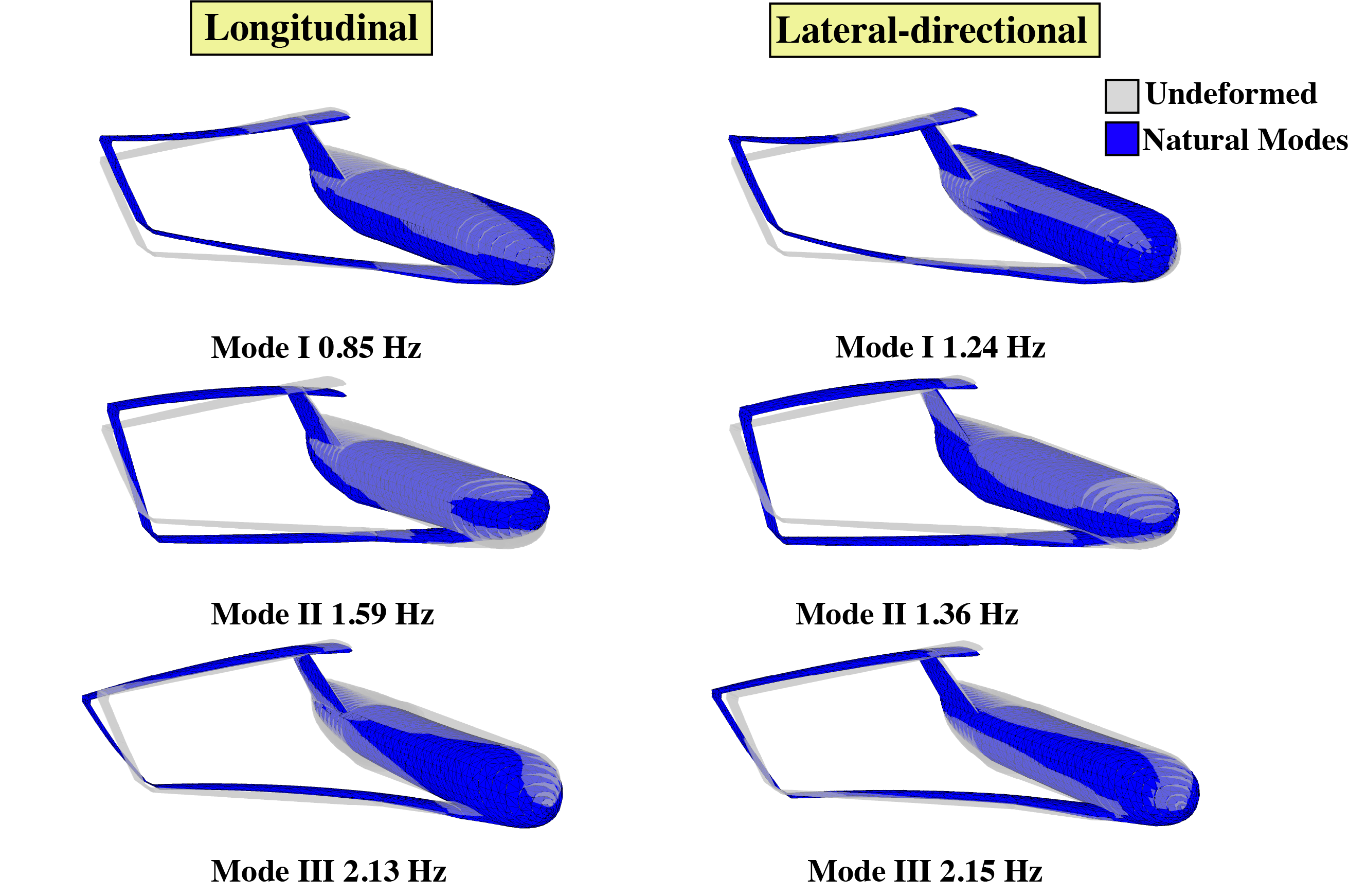}
	\caption{Artistic view of the first three longitudinal (left) and lateral-directional (right) natural modes of the free-flying PrP250, and relative frequencies.}
	\label{f:PrP_modes}
\end{figure}

Modal properties of the computational model, in terms of shape and frequency of the first three natural modes, are summarized in Fig.~\ref{f:PrP_modes}. 
Due to the overconstrained nature of this characteristic wing system layout, bending/torsion coupling of both wings is inherently enhanced. 
The \emph{first} mode is characterized by a vertical deflection of the wings, with an (almost) rigid vertical translation of the vertical joint. In the \emph{second} mode the joint tilting dominates the deformation, inducing wing bending and torsion. Moreover, the joint itself slightly bends. 
On the contrary, in the \emph{third} mode the joint (almost) rigidly translates and slightly tilts inward/outward~\cite{SDSUteam_6jour}. 
The model is free in the space and fuselage motion is observed mainly as pitching and plunging in the longitudinal modes, and as rolling and yawing, with negligible side motion, in the lateral-directional ones.
%
%
%
%
%
%
\par
\begin{figure}[htb]
	\centering
	\vspace{5mm}
	\includegraphics[width=\textwidth]{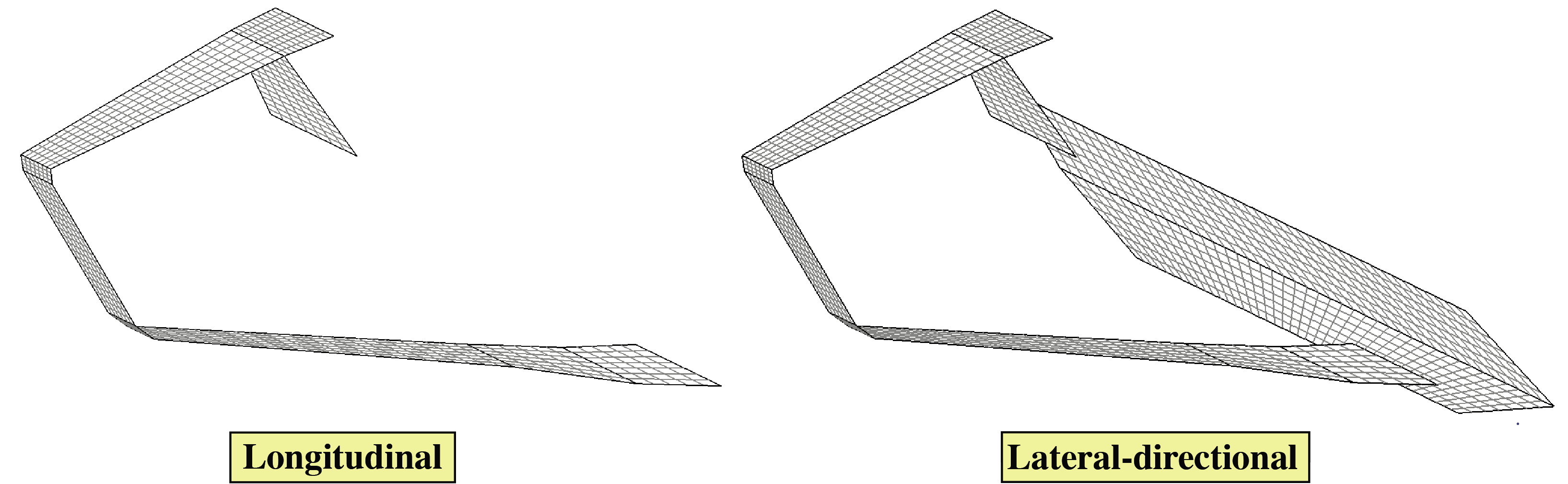}
	\caption{Aerodynamic model for longitudinal (left) and lateral-directional (right) analyses.}
	\label{f:PrP_aero}
\end{figure}

 
The aerodynamic mesh used in the DLM is shown in Fig.~\ref{f:PrP_aero}; it consists of approximately 800 panels for the longitudinal model and 1500 for the lateral-directional one, and proved to give converged results. 
%
%
%
%
%
%
%
%
An overview of the model inertial parameters, with respect to the reference system shown in Fig.~\ref{f:PrP_fem}, is provided in Table~\ref{t:Free_free_inertial}. The reference system is the DLM one, with the $x$-axis along the fuselage and $y$-axis along the wing span. 
%
%
\begin{table}[!ht]
	\centering
	\begin{tabular}{@{}lcrc@{}}
		\toprule
		Mass (MTOW) & $M^{\text{ref}}$ & $230 \cdot 10^3$ & kg \\
		Rolling moment of inertia & $I_x^{\text{ref}}$ & $1.50 \cdot 10^7$ & kg $\cdot$ m$^2$ \\		
		Pitching moment of inertia & $I_y^{\text{ref}}$ & $3.53 \cdot 10^7$ & kg $\cdot$ m$^2$ \\
		Yawing moment of inertia & $I_z^{\text{ref}}$ & $4.44 \cdot 10^7$ & kg $\cdot$ m$^2$ \\ \bottomrule		
	\end{tabular}
	\caption{Inertial parameters. Moments of inertia are evaluated about the DLM reference system shown in Figure~\ref{f:PrP_fem}(b).}
	\label{t:Free_free_inertial}
\end{table}
%
\section{Results}      %
%
%
%
In this section, results of the investigation relative to the response of the free-flying PrP250 aircraft are shown. First, flying qualities are discussed, for both the longitudinal and lateral-directional dynamics.
Then, the aeroelastic viewpoint is adopted. 
Results for both longitudinal and lateral-directional dynamics are presented, and focus is on the physical sources of differences between  the case in which the aircraft is fixed or free in the space. 
The section ends with a discussion on the effects of the enhanced DLM on flight-dynamic and aeroelastic response.
%
%
%

%
%
%
\subsection{Flight conditions and settings}
For this study, the aircraft is considered at straight-and-leveled flight. 
Two conditions in the flight envelope are considered. Both refer to the aircraft at $V_C$ (design cruise speed), but at the different altitudes, namely, cruise altitude $h_c = 10500$~m and sea level. 
At nominal cruise altitude Mach is $0.85$ for a true airspeed of $V_\infty=252$~m/s. At sea level,  airspeed is $V_\infty=196$~m/s (equivalent to a Mach number of approximately  $M=0.58$). 
%
\par
%
Compressibility correction, based on the Prandtl-Glauert rule, is used for aerodynamic calculations
for both considered altitudes. 
\par
For the analyses, in the case of flexible aircraft 25 normal modes, i.e., three rigid and 22 elastic for each case (longitudinal and lateral-directional) have proven to ensure convergence of results.
%
GAFs are calculated for a set of 21 reduced frequencies, adequately chosen to describe the physics of interest.
RFA process consists in a modified Roger method, using 6 lag terms, to resolve accurately the dynamics for frequencies typical of both flight dynamics and aeroelastic phenomena, as described in section~\ref{sss:rational}.
%
%
%
\par%
In the case of rigid aircraft, the three rigid modes are trivially selected. A quasi-steady approximation has been conveniently performed to correctly model Short Period and Dutch Roll.
%
%
%
%
\par
It is stressed out that the enhanced DLM requires the evaluation of the aerodynamic flow properties (such as AoA and vortex circulation distribution) in the reference flight conditions.
Hence, for each flight condition, a preliminary aerodynamic analysis is carried out by means of the VLM. 
Moreover, when using the enhanced DLM, the AoA of the aircraft in the sought flight condition influences the rotation of the vehicle reference moments of inertia (matrix $\bM_{RR}$ of Eq.~(\ref{eq:rigid_matrix_salf2_long}) specialized to the lateral-directional dynamics).
%
%
\par
When, for purpose of comparison, the aircraft has been considered fixed in the space, the structural model has been clamped at its center of mass, and symmetric or anti-symmetric constraints have been applied on the symmetry plane, in accordance with the analysis to perform. A traditional aeroelastic analysis has been carried out considering 22 elastic modes, and using the enhanced DLM; moreover, RFA has been performed using 6 lag terms on
the GAFs calculated on the 21 reduced frequencies
covering the physical phenomenon to be caught. 
%
\subsection{Flight dynamics of the flexible PrP}   
%
To assess the aircraft handling qualities, requirements provided by the MIL-F-8785C~\cite{SpecificationMILF} are referenced for the cases of Short Period and Dutch Roll. Such requirements relate handling qualities, ranging from 1 (best) to 3 (sufficient), to the eigenvalues properties such as damping ratio and/or frequency. 
%
Prescribed requirements are specific for the vehicle class and chosen flight phase. 
The PrP250 is classified as aircraft of class III (``large, heavy, low-to-medium maneuverability airplane''). 
Only cruise condition is considered within this context, which is included in category B.
%
\par
For both longitudinal and lateral-directional dynamics, flying qualities for rigid and flexible aircraft are reported and compared in order to assess the effects of flexibility.
%
\subsubsection{Short Period and flying qualities}  
\begin{figure}[htb]
	\centering
	\includegraphics[width=\textwidth]{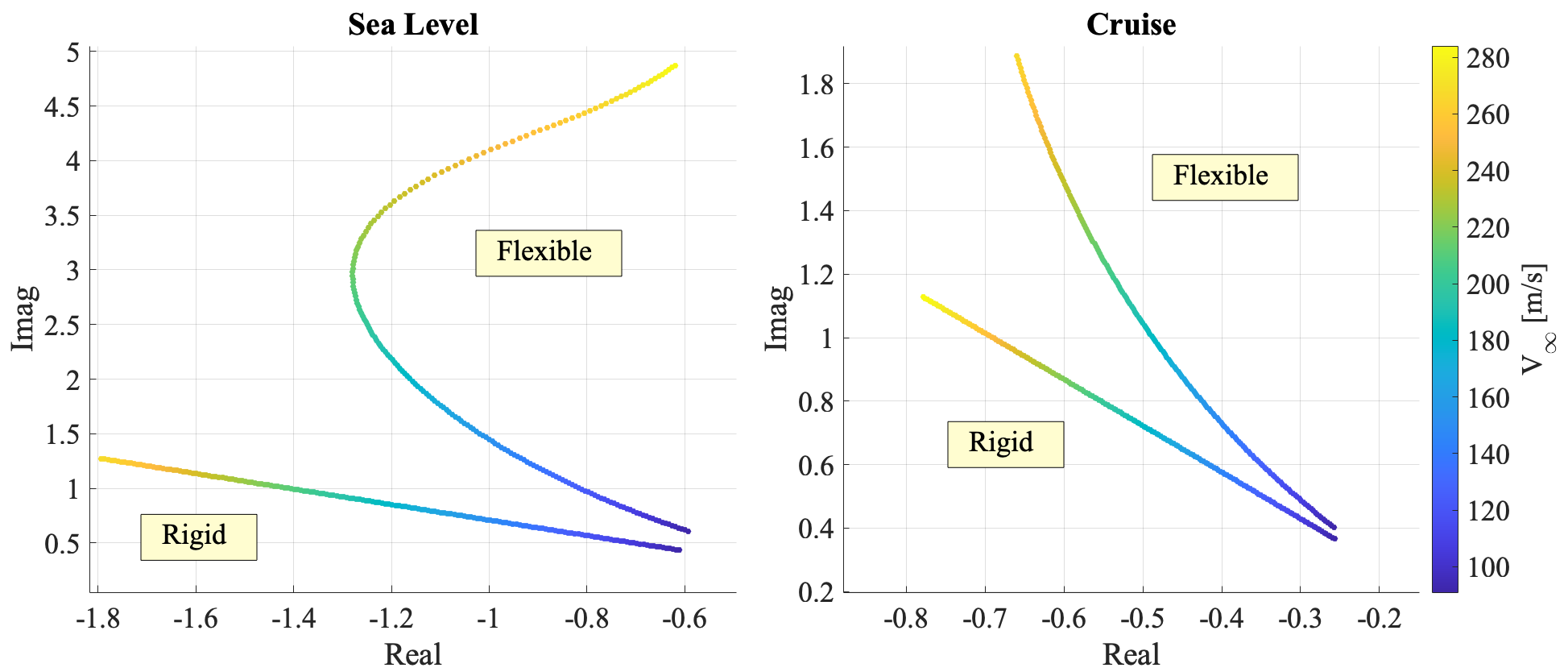}
	\caption{Comparison of Short Period eigenvalues in the complex plane for rigid and flexible PrP250 at sea and cruise levels.}
	\label{f:Short_period_comparison}
\end{figure}

%
The comparison of the Short Period for flexible and rigid vehicle at sea and cruise levels is shown in Fig.~\ref{f:Short_period_comparison}.
%
%
It can be noted how, as speed increases the flexible branch moves towards the imaginary axis and at higher frequencies, for both cruise conditions. It will be shown in the following section how this behaviour is consequence of the aerodynamic interaction with elastic modes.
\par
%
%
For the Short Period the MIL-F-8785C prescribes specific values on the damping ratio, as shown in Tab.~\ref{t:MIL_Short_period} for a class III aircraft.
%
%
%
\begin{table}[!t]
\centering
\begin{tabular}{c c}
\hline\noalign{\smallskip}
 \multicolumn{1}{l}{\textbf{Requirements}}  &  \\ 
\noalign{\smallskip}\hline\noalign{\smallskip}
 \textbf{Flying Quality Level} & \textbf{Flight Phase Category B}     \\ 
 \noalign{\smallskip}\hline\noalign{\smallskip}
  1 & $0.30 \leq \zeta_{SP} \leq 2.00$\\
  2 & $0.20 \leq \zeta_{SP} \leq 2.00$\\
  3 & $0.15 \leq \zeta_{SP} $\\
\noalign{\smallskip}\hline
\end{tabular}
\caption{Flying quality requirements for Short Period damping ratio: aircraft of class III during cruise (Category B).}
\label{t:MIL_Short_period} 
\end{table}
%
Tab.\ref{t:short_period_PrP250} summarizes the values of damping ratio found for rigid and flexible aircraft.  
For the sake of clarity, only values of the poles for $V_C$ at both considered altitudes are shown.
%
Comparing data in Tab.~\ref{t:short_period_PrP250} with the requirements in Tab.~\ref{t:MIL_Short_period} it can be inferred that the considered configuration  fulfills the requirements for Flying quality level 1,  both for the rigid and the flexible cases. 
It is worth to underline, however, the significant reduction of $\zeta_{SP}$ when flexibility of the aircraft is considered: namely, a $40\%$ and $32\%$ reduction at sea level and cruise altitude, respectively, which pushes the flying quality towards level 2. 
This demonstrates how structural flexibility of the aircraft is a crucial aspect to consider when studying the PrP250 flying qualities.
%
%
\begin{table}[!t]
\centering
\begin{tabular}{ c c c c c }
\hline\noalign{\smallskip}
&  \multicolumn{2}{c}{\textbf{Sea level}} &  \multicolumn{2}{c}{\textbf{Cruise}} \\
\noalign{\smallskip}\hline\noalign{\smallskip}
& \textbf{Rigid} & \textbf{Flexible} & \textbf{Rigid} & \textbf{Flexible} \\
Eigenvalue & -1.31 + i0.93 & -1.23 + i2.40 & -0.69 + i1.00& -0.63 + i1.69\\
$\omega_{n_{SP}} [rad/s]$ &1.6 & 2.69& 1.21  & 1.80 \\
$ \zeta_{SP}$         &0.82   & 0.45  & 0.57  &  0.35\\
\noalign{\smallskip}\hline
\end{tabular}
\caption{Short Period eigenvalues at sea and cruise level for the PrP250.}
\label{t:short_period_PrP250} 
\end{table}
%
\subsubsection{Dutch Roll and flying qualities}  
\begin{figure}[htb]
	\centering
	\includegraphics[width=\textwidth]{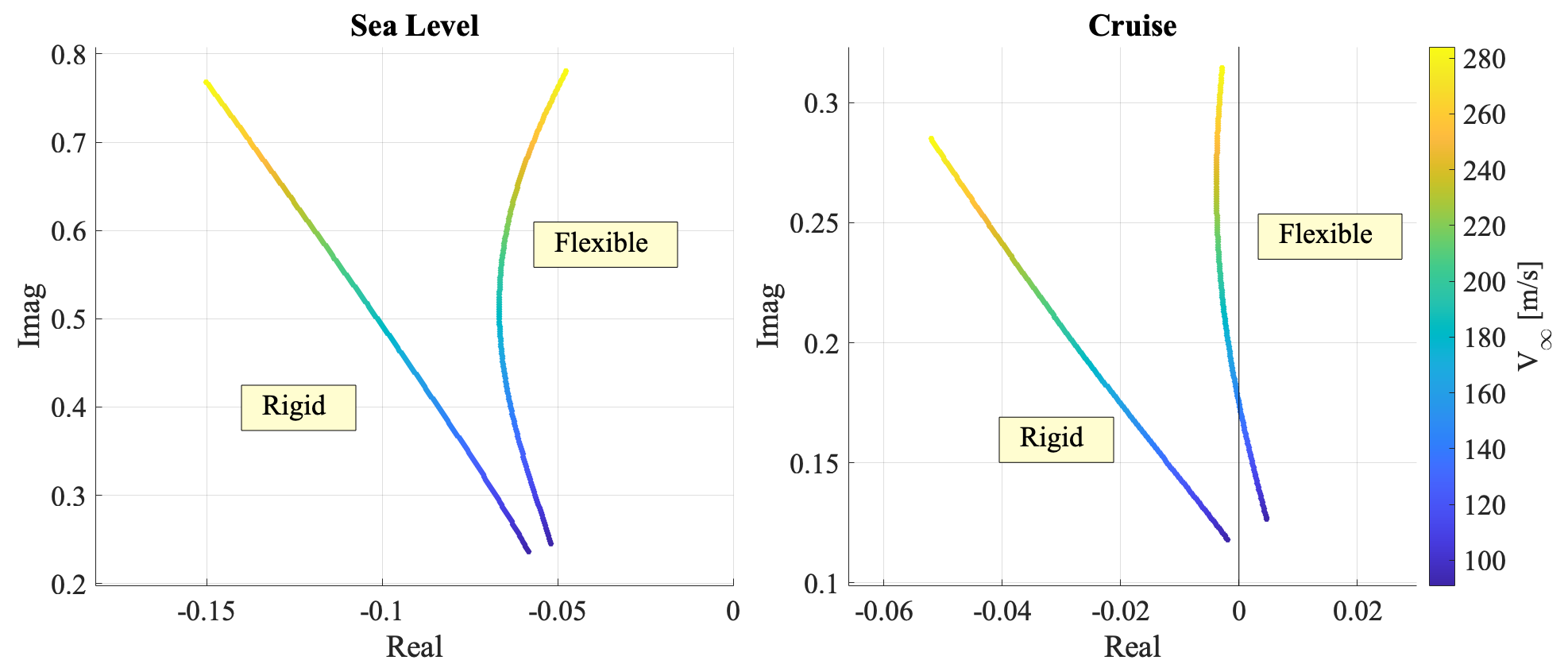}
	\caption{Comparison of Dutch roll eigenvalues in the complex plane for rigid and flexible PrP250 at sea and cruise level.}
	\label{f:Dutch_roll_comparison}
\end{figure}

%
The comparison of the Dutch Roll mode for flexible and rigid vehicle at sea and cruise level is shown in Fig.~\ref{f:Dutch_roll_comparison}. 
It can be appreciated how considering the elastic interaction shifts the branch associated to the mode towards the imaginary axis, similarly to what seen for the Short Period. 
For cruise altitude an flexible aircraft Dutch Roll is unstable at low speeds, becoming stable at a speed of approximately $130~m/s$. 
It must be stressed out, however, that these analyses have been carried out linearizing about $V_C$, hence, 
while results may be considered quantitatively valid around the reference condition, they only retain qualitative meaning far from them.
%
%
%
%
%
\par
Requirements provided by the MIL-F-8785C for the Dutch Roll mode are based on minimum values of the damping ratio, frequency and their product (absolute value of the real part of the eigenvalue) assuming a stable Dutch Roll, as shown in Tab.~\ref{t:MILF_lateral}. 
%
\begin{table}[!ht]
	\centering
	\begin{tabular}{ccccc}
		\hline\noalign{\smallskip}
		 \multicolumn{2}{l}{\textbf{Requirements}}  &  \\ 
		\noalign{\smallskip}\hline\noalign{\smallskip}
		\textbf{Flying Quality Level} & \begin{tabular}[c]{@{}c@{}}\textbf{Flight Phase}\\ \textbf{Category}\end{tabular} & \multicolumn{1}{c}{\textbf{Min $\zeta_{D}$}} & \multicolumn{1}{c}{\begin{tabular}[c]{@{}c@{}}\textbf{Min $\zeta_{D} \, \omega_{n_D}$}\\ \textbf{{[}rad/s{]}}\end{tabular}} & \multicolumn{1}{c}{\begin{tabular}[c]{@{}c@{}}\textbf{Min $\omega_{n_D}$}\\ \textbf{{[}rad/s{]}}\end{tabular}} \\
		\noalign{\smallskip}\hline\noalign{\smallskip}
		\multirow{3}{*}{1} & A & 0.19 & 0.35 & 0.4 \\
		& B & 0.08 & 0.15 & 0.4 \\
		& C & 0.08 & 0.10 & 0.4 \\ 
		\noalign{\smallskip}\hline\noalign{\smallskip}
		2 & All & 0.02 & 0.05 & 0.4 \\ 
		3 & All & 0.00 & 0.00 & 0.4 \\ 
		\hline\noalign{\smallskip}
	\end{tabular}
	\caption{Flying quality level requirements for Dutch Roll according to the MIL-F-8785C for vehicles of class III.}
	\label{t:MILF_lateral}
\end{table}
%
Tab.~\ref{t:dutch_roll_PrP250} shows the Dutch Roll eigenvalues for the rigid and flexible cases at $V_C$, both at sea level and cruise altitude. 
It can be inferred  that the rigid aircraft flying qualities are level 2 at sea level, whereas, at the nominal cruise altitude the values of $\omega_{n_D}$ are not even compatible with level 3. 
%
When flexibility of the structure is taken into account, a noticeable drop in the  eigenvalue damping ratio is observed: a $40\%$ and a $92\%$ reduction at sea and cruise level, respectively. 
At sea level, switching from rigid to flexible configuration, the flying quality almost downgrades from level 2 to level 3. 
%
\begin{table}[!t]
\centering
\begin{tabular}{ c c c c c }
\hline\noalign{\smallskip}
&  \multicolumn{2}{c}{\textbf{Sea level}} &  \multicolumn{1}{c}{\textbf{Cruise}} \\
\noalign{\smallskip}\hline\noalign{\smallskip}
& \textbf{Rigid} & \textbf{Flexible} & \textbf{Rigid} & \textbf{Flexible} \\
Eigenvalue                    & -0.11 + i0.52 & -0.06 + i0.54 & -0.04 + i0.26& -0.00 + i0.28\\
$\omega_{n_{D}} [rad/s]$     &0.54           & 0.55              & 0.25             & 0.28 \\
$ \zeta_{D}$                  &0.20         & 0.12            & 0.17             &  0.01\\
\noalign{\smallskip}\hline
\end{tabular}
\caption{Dutch Roll eigenvalues at sea and cruise levels for the PrP250.}
\label{t:dutch_roll_PrP250}   
\end{table}
%
\par
%
%
%
It is a known fact that reliable computational prediction of the aerodynamic derivatives is  cumbersome. Moreover, a research having the ambition of assessing with high level of accuracy the flying qualities of the herein investigated unconventional configuration would require data such as a more precise model for the inertia of the vehicle (e.g., fuselage and non-structural masses) and a detailed aerodynamic model of the whole aircraft. 
Flying qualities level requirements, here used, serve as a criterion for comparison, and Short Period and Dutch Roll eigenvalues are benchmarks to assess the effects of structural flexibility on stability, which is one of the sought contributions of this work.
%
\subsection{Aeroelasticity}    %
%
%
In the previous section focus was on flight dynamics, thus, effects of the flexibility of the structure in terms of flying qualities were discussed. 
A broader perspective can be earned recalling that the unified  analyses  presented in the previous  section  for  the  flexible  aircraft  give as outputs all the eigenvalues of the system and allow to identify the effects of aerodynamic interaction between the flight-dynamic and the aeroelastic modes. 
%
Hence, it is possible to observe and explain the stability of the free-flying aircraft with respect to the fixed-in-space configuration.
%
\par
These changes in the response are driven by two physical reasons. First, when the aircraft is free in space, the elastic normal modes are (slightly) different in shape and frequency than their counterparts for the fixed-in-space case, due to the different structural boundary conditions of the eigenvalue problem. 
As modes and frequencies are different, also aeroelastic response is expected to be different. 
The second source of difference comes from the aerodynamic interaction between the rigid and elastic modes. 
With reference to the theoretical section~\ref{s:FDDA}, it has been shown how the aerodynamic operator is the only one carrying all rigid/elastic coupling effects, i.e., the equations regulating the rigid and elastic Lagrangian coordinates response are coupled only by means of the aerodynamic operator. 
%
%
%
\par
%
%
%
%
%
%
%
\subsubsection{Longitudinal case}
\begin{figure}[htb]
	\centering
	\includegraphics[width=\textwidth]{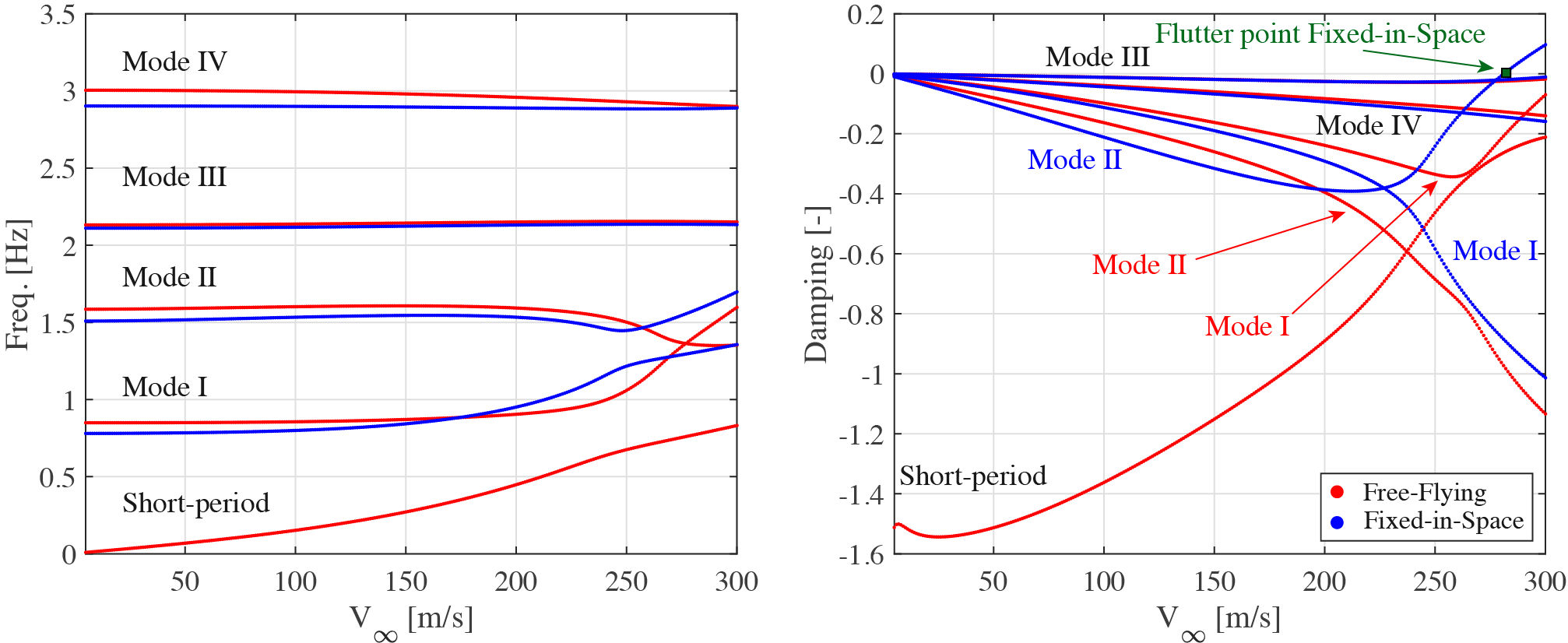}
	\caption{PrP250 stability diagram for the longitudinal dynamics: frequency and damping (at sea level). Comparison between fixed-in-space and free-flying configurations.}
	\label{f:Flutter_long}
\end{figure}

%
%
Results of the longitudinal stability analysis performed with UFFD on the PrP250, in terms of frequency and damping coefficients of the system eigenvalues versus the asymptotic speed, are shown in Fig.~\ref{f:Flutter_long} at sea level conditions. 
%
%
As a comparison, results of the stability analysis for the fixed-in-space configuration are given in the same figure.
%
A similar comparison was already proposed in ~\cite{SDSUteam_6jour} for the PrP250, although that study was a pioneering one, neglecting effects of compressibility, formulating the perturbation equations without considering flight dynamics nor the effects of the reference condition, and using a classic DLM as aerodynamic prediction tool. 
%
%
%
%
%
The comparison is able to highlight the different behaviors induced by the presence of the rigid modes on the flutter response. 
%
%
As seen from Fig.~\ref{f:Flutter_long}, for the fixed-in-space configuration flutter occurs at approximately 281 m/s while the free-flying configuration is flutter free in the considered speed range, up to 300 m/s. 
%
%
In work~\cite{SDSUteam_6jour} an interpretation of such phenomenon was suggested for which an interaction between rigid and elastic modes was postponing flutter onset.  
\par
\begin{figure}[htb]
	\centering
	\includegraphics[width=\textwidth]{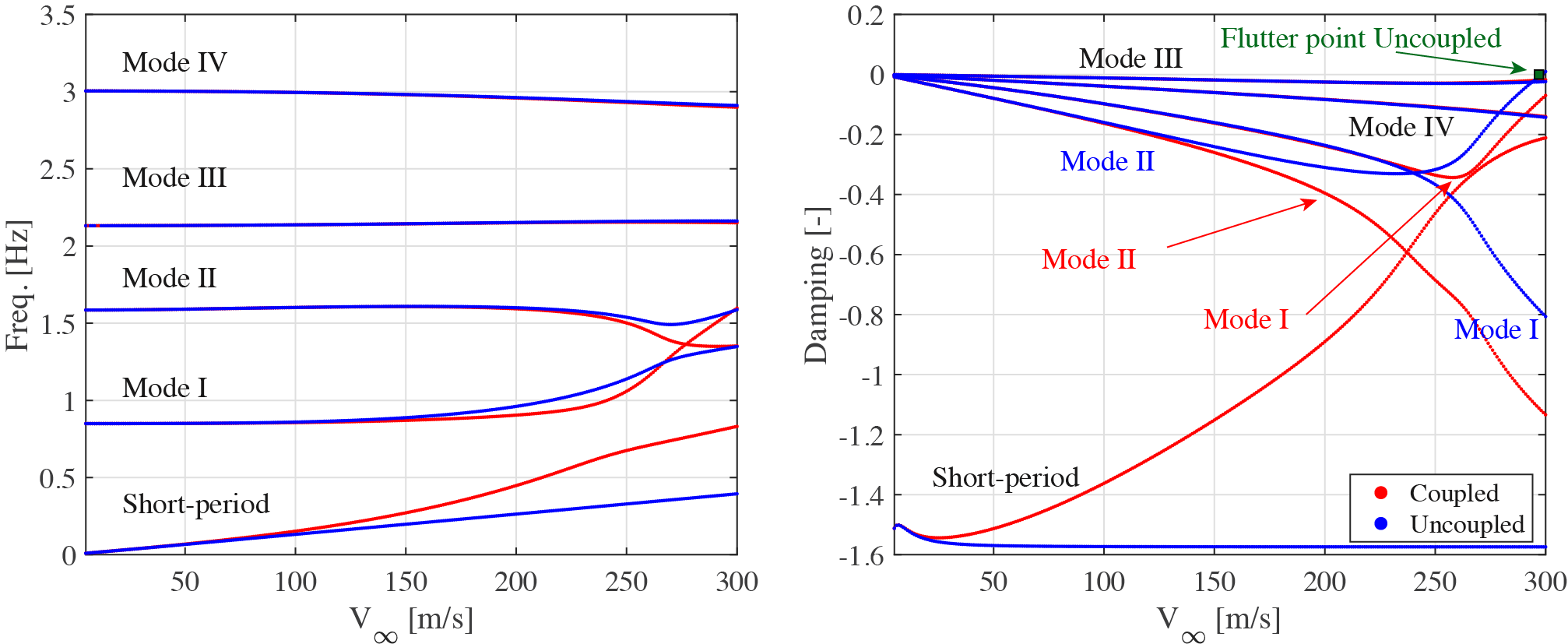}
	\caption{PrP250 free-flying stability diagram for the longitudinal dynamics: frequency and damping of the system at  sea level when considering and neglecting aerodynamic coupling.}
	\label{f:Flutter_long_uncoupled}
\end{figure}

In order to endorse such interpretation, a comparison is carried out on the free-flying aircraft, including or neglecting the aerodynamic coupling between equations governing the rigid and elastic generalized coordinates dynamics: i.e.,
%
%
selectively setting $\bA_{i_{RE}}$ and $\bA_{i_{ER}}$ to zero (as shown in Eq.~\ref{eq:aero_coupling} in section~\ref{sss:flexible_aerodynamics}).
%
Results of such comparison is shown in Fig.~\ref{f:Flutter_long_uncoupled}.
%
%
It can be seen how, for the uncoupled analysis, the Short Period mode doesn't interact with the elastic modes, as expected  and differently than the coupled analysis. 
Such interaction is beneficial as it raises flutter speed. 
%
Nevertheless, flutter speed is still different from the one of the fixed-in-space configuration. 
This is due to the fact that the two considered configurations exhibit different free vibration properties (shape and frequency of structural modes). 
It can be concluded then, that the raise in flutter speed observed for the free-flying configuration is due to the synergistic effect of the change in free vibration modes and the aerodynamic interaction with flight-dynamic modes, being the second contribution the dominant one.
%
%
%
\subsubsection{Lateral-directional case}    %
\begin{figure}[htb]
	\centering
	\includegraphics[width=\textwidth]{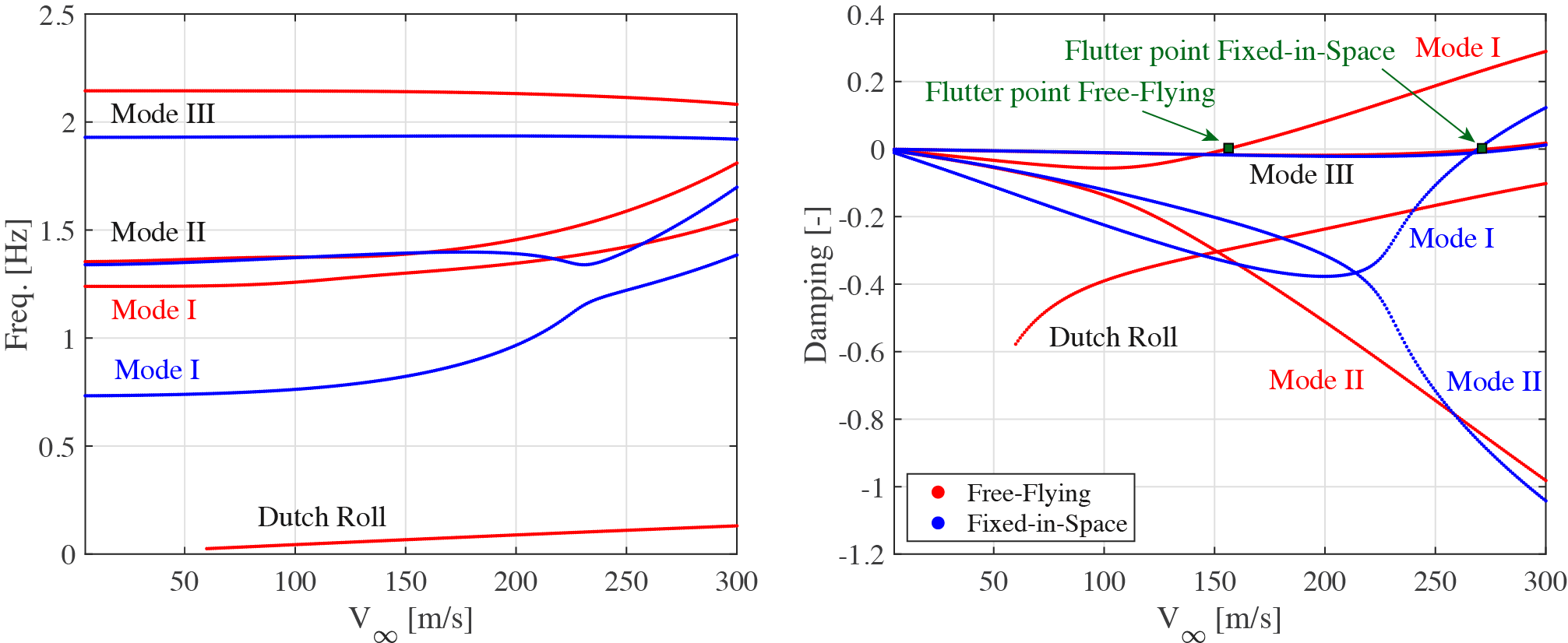}
	\caption{PrP250 stability diagram for the lateral-directional dynamics: frequency and damping of the system at sea level. Comparison between fixed-in-space and free-flying configurations.}
	\label{f:Flutter_lateral_comp}
\end{figure}

%
%
Fig.~\ref{f:Flutter_lateral_comp} depicts the lateral-directional stability analysis performed with UFFD in terms of frequency and damping coefficients of the system eigenvalues. 
%
%
%
Similarly to what done for the longitudinal case, the unified analysis of the free-flying configuration is compared with the flutter results of the fixed-in-space system. 
Trends are  diverse between the two cases, resulting in completely different
flutter speeds: whereas   for the fixed-in-space configuration the flutter onset is at approximately 270 m/s, for the free-flying one it drops to 155 m/s. In both cases the first elastic mode becomes unstable.
%
%
\par
It is interesting to find out what is the main source of this large difference. 
Observing the first elastic mode in both cases, it can be observed how its natural frequency is larger for the free-flying configuration, and gets very close to the frequency of the second elastic mode, possibly favouring a mutual interaction and driving the early onset of the instability. 
%
\begin{figure}[htb]
	\centering
	\includegraphics[width=\textwidth]{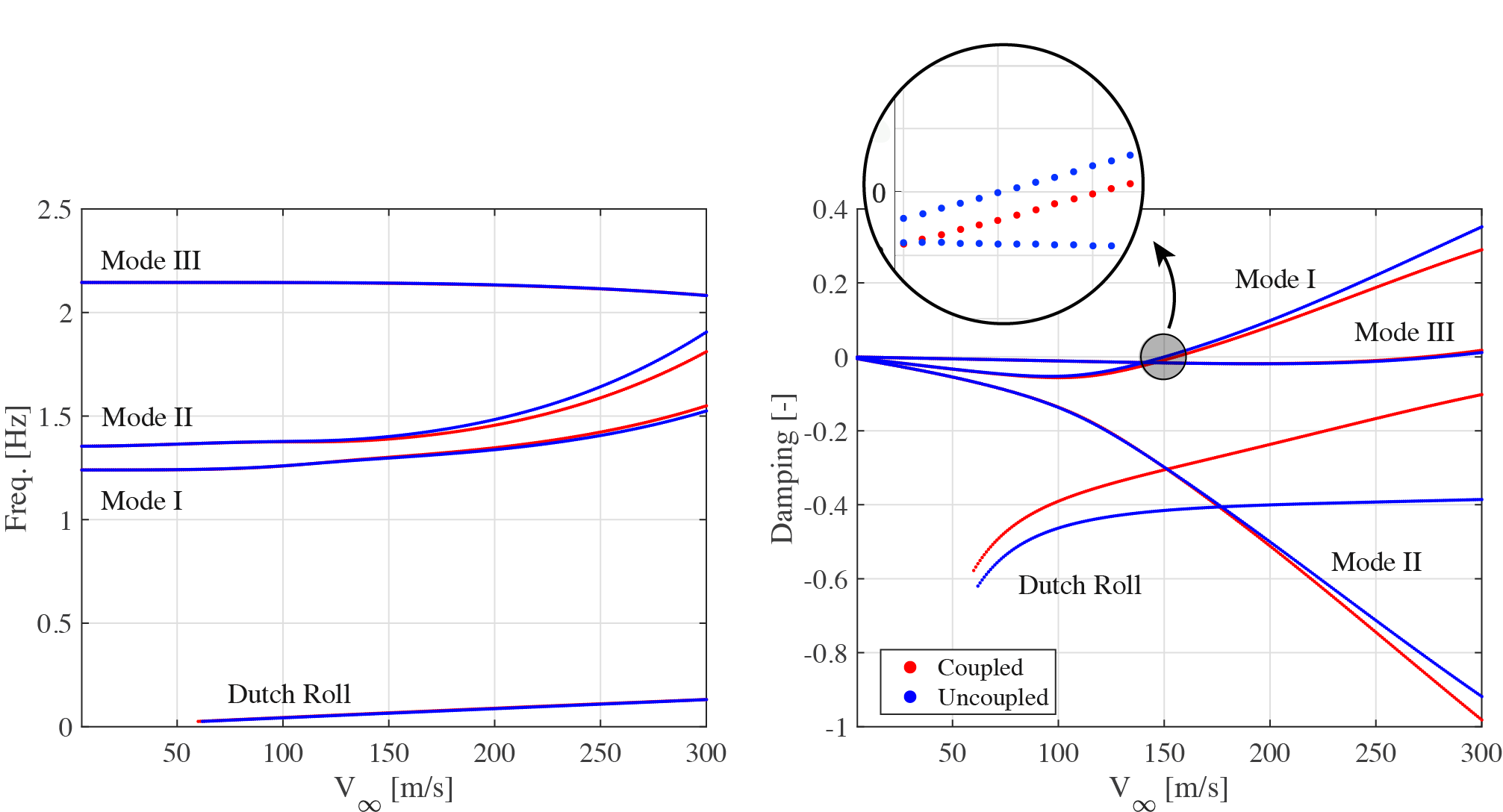}
	\caption{PrP250 free-flying stability diagram for the lateral-directional dynamics: frequency and damping of the system at  sea level when considering and neglecting aerodynamic coupling.}
	\label{f:Flutter_lateral_uncoupled}
\end{figure}
%
%
%
%
%
For a more in-depth understanding, same as what already done for the longitudinal case, a comparison is carried out between cases in which aerodynamic coupling
is neglected or taken into account. 
\par
Result of such comparison is shown in Fig.~\ref{f:Flutter_lateral_uncoupled}. 
It can be inferred that flutter speed slightly increases, as effect of the aerodynamic interaction between aeroelastic modes and Dutch Roll. 
%
In view of the here-shown results, it can be argued that the low flutter speed does not depend on a detrimental aerodynamic interaction between rigid and elastic modes, but, simply on the different shape and frequency that elastic modes have with respect to the fixed-in-space model. 
%
%
\subsection{Effect of DLM corrections}    %
\begin{figure}[htb]
	\centering
	\includegraphics[width=\textwidth]{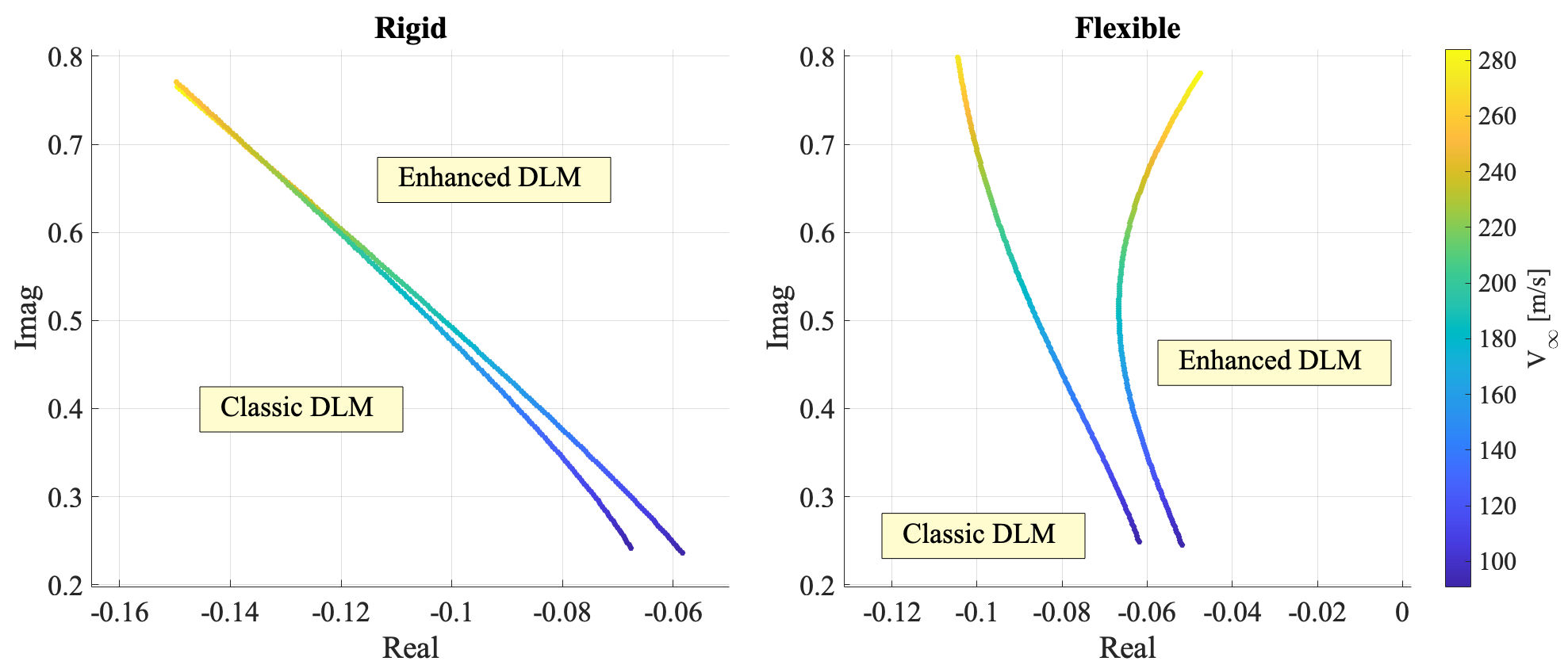}
	\caption{Comparison of Dutch Roll eigenvalues evaluated with classic and enhanced DLM for the rigid and flexible vehicle at sea level for the PrP250.}
	\label{f:Dutch_roll_enhancedDLM}
\end{figure}

%
The above results have been obtained considering the enhanced DLM, which, with respect to the classic formulation, features a more general aerodynamic boundary condition able to take into account yaw-dihedral coupling, and employs a force calculation method which considers, at aerodynamic panel level, lateral loads and yawing moments due to roll rate and rolling moments due to yaw and yaw rate (see section \ref{s:enhancedDLM}). 
These contributions have been observed to be negligible for the longitudinal behavior, however, they have a relevant impact on the lateral-directional dynamics.
%
%
Fig.~\ref{f:Dutch_roll_enhancedDLM} compares the Dutch Roll for the rigid and flexible configurations, at sea level, when the classic and the enhanced DLM are employed. 
%
%
%
%
\par
%
%
It can be clearly seen how, for this aircraft configuration and for both the rigid and the flexible cases, the tendency of a more inclusive generalized force calculation procedure is to push the Dutch Roll mode at lower frequency and absolute value of the eigenvalue real part. Moreover, with the enhanced DLM effects of flexibility seem to be even more relevant. 
\par
\begin{figure}[htb]
	\centering
	\includegraphics[width=\textwidth]{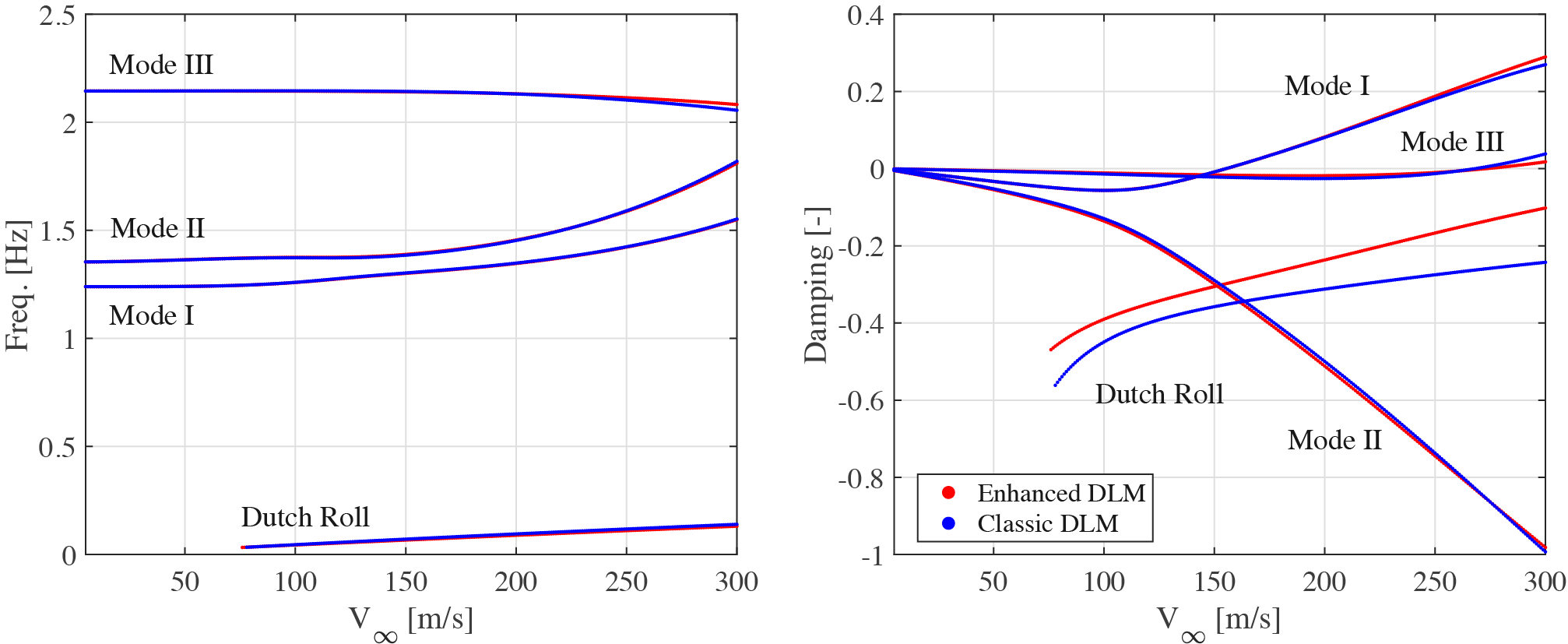}
	\caption{PrP250 free-flying stability diagram for the lateral-directional dynamics: frequency and damping of the system at sea level when considering enhanced or classic DLM.}
	\label{f:Flutter_lateral_enhanced_DLM}
\end{figure}

When focusing on aeroelasticity, effects of extra terms are not relevant, as shown in Fig.~\ref{f:Flutter_lateral_enhanced_DLM}. 
It is possible to speculate that the PrP250 configuration, with its particular distribution of stiffness and  with its double fin configuration, avoids physical instability phenomena typical of T-tail configuration, that can only be modeled with such extra terms~\cite{VanZyl-2011}. 
%
%
%
%
\par
A quantitative comparison of the differences when selectively including the extra terms into the DLM is shown in Tab.~\ref{t:enhanced_DLM coefficient} for some GAF entries  relative to the rigid modes, namely $\widetilde{\bA}_{0_{RR}}$ and $\widetilde{\bA}_{1_{RR}}$.  
In this case a quasi-steady approximation has been chosen; 
in such way the selected GAF have a clear relation with the aerodynamic derivatives, aiding for a physical interpretation~\cite{UC3M_UFFDScitech}. 
%
%
\begin{table}[!ht]
	\centering
	\begin{tabular}{lccccc}
		\hline\noalign{\smallskip}
		  & \textbf{Classic} & \textbf{Boundary Corr.} & \textbf{1st Term Corr.} &  \textbf{3rd Term Corr.} &  \textbf{All} \\
		\noalign{\smallskip}\hline\noalign{\smallskip}
		$\widetilde{\bA}_{0}^{13}$ &-108.0&-107.8&-100.2&-108.0&-100.0\\
		$\widetilde{\bA}_{0}^{23}$ &-228.8&-228.5&-259.2&-228.8&-258.7\\
		$\widetilde{\bA}_{0}^{33}$ &-329.8&-329.2&-290.6&-329.8&-290.1\\
		$\widetilde{\bA}_{1}^{12}$ &476&512&215&276&47\\
		$\widetilde{\bA}_{1}^{22}$ &-130550&-130722&-132207&-130550&-132354\\
		\noalign{\smallskip}\hline
\end{tabular}
\caption{Effects of terms of the enhanced DLM on some entries of $\widetilde{\bA}_{0_{RR}}$ and   $\widetilde{\bA}_{1_{RR}}$. Dimensions are [$m^2$]. $RR$ subscript omitted for clarity. Indices are: 1 for side slip, 2 for roll, 3 for yaw.}
\label{t:enhanced_DLM coefficient} 
\end{table}
%
For example, observe the change in the term $\widetilde{\bA}_{0_{RR}}^{23}$, which has a strong correlation with the rolling moment coefficient due to sideslip $C_{l_{\beta}}$, or the variation in the term $\widetilde{\bA}_{1_{RR}}^{12}$, related to the side force coefficient due to roll rate $C_{Y_{p}}$.
%
%
GAFs relative to elastic modes also change due to correction terms, however, the final effect on the flutter onset is of second order.
%
%
%
%
%
%
%
\par
It is important to stress out that, considering quasi-steady aerodynamics only, VLM can be used to predict all but unsteady GAF terms, existing a clear relation between aerodynamic matrices as in Eq.~(\ref{eq:faero_FT2}) and aerodynamic stability derivatives~\cite{Baldelli_JoA2006,UC3M_UFFDScitech}.
However, the relevance of using the DLM lies in its ability of predicting unsteady aerodynamic terms and hence, of conveniently reproducing the physics of interest, which covers now both low and high frequencies.
\par
In reference~\cite{Baldelli_JoA2006} the possibility is discussed to selectively correct aerodynamic matrices, calculated by means of the chosen RFA method, using different methods (see section~\ref{sec:introduction_UFFD}).
However, this is, consistently with the RFA strategy used within this investigation, only possible when choosing a quasi-steady aerodynamic approximation.
In the unified case, when a broad range of frequencies needs to be resolved, the quasi-steady approximation is not viable anymore, and, after performing RFA, it is not possible to selectively correct a-posteriori the aerodynamic matrices, as there is not a one-to-one correspondence with the aerodynamic derivatives. 
Hence, correction of the aerodynamic coefficients provided by the traditional DLM cannot be performed easily: from here the relevance of adopting an enhanced DLM. 
It should be mentioned, however, that some literature efforts have proposed methods levering on a different form of RFAs to allow selective corrections of the GAFs \cite{Kier2011AnIL}, although at the price of a largely increased number of added states. 
%
%
%
%
%
%

%
\FloatBarrier
\section{Conclusions}      %
%
In this study, the application of a unified flight-dynamic and aeroelastic framework for the stability analysis of a PrandtlPlane configuration is discussed. 
The investigation covers both longitudinal and lateral-directional dynamics and aims at assessing the mutual interaction between the two disciplines on the aircraft stability. 
\par 
First, the formulation is derived and discussed. An enhanced DLM method is used for aerodynamic forces evaluation and a Roger-based Rational Function Approximation strategy is chosen to interpolate such forces over the range of frequencies of interest.
\par
Second, flying qualities for Short Period and Dutch Roll modes are assessed and discussed for both the rigid and flexible configurations. 
A remarkable degradation in terms of damping ratio of the relative poles is noted when considering the flexible configuration, as opposed to the rigid one, for both the longitudinal and lateral-directional cases. 
Results remark the importance of considering the vehicle elasticity when assessing its flying qualities. 
\par
Third, with respect to aeroelasticity, results shed light on phenomena preliminary observed in a few previous literature efforts, for which flutter speed were notably different when analyzing the fixed-in-space or free-in-the-air PrandtlPlane.
Thanks to the possibility, in the equations governing the dynamics of the flexible and free-flying aircraft, of selectively including rigid/elastic aerodynamic coupling effects, the source of changes in flutter speed between the two cases has been explained.
With respect to the longitudinal case, the free-flying aircraft has a flutter speed significantly larger than the fixed-in-space one. 
This is consequence of the synergistic effect of the aerodynamic interaction between rigid and elastic modes and the change in normal modes due to different structural boundary conditions, being the first effect the dominant one. 
%
%
\par
For the lateral-directional case, the free-flying configuration experiences a considerable drop in flutter speed, well inside the flight envelope. 
Also in this case the aerodynamic coupling proves to give a benign effect, even though small, with a tendency of increasing flutter speed.
However, the change in the  normal modes due to the different boundary conditions has a dominant detrimental effect, inducing an early flutter onset. 
%
%
\par
%
Finally, effects of the enhanced DLM are assessed, showing how, for this configuration, it changes mainly the flight-dynamic response, and exacerbates effects of flexibility. From an aeroelastic perspective, even though GAFs values do change, no relevant effects are registered on stability properties. 

\section*{Declaration of competing interest}
%
The authors declare that they have no known competing financial interests or personal relationships that could have appeared to influence the work reported in this paper.
%
\section*{Acknowledgements}
%
\bibliographystyle{unsrt}

\bibliography{./bibliography/Bibliography_UFFD}
%
\end{document}